\documentclass{aa}  

\usepackage{caption}
\usepackage{subcaption}

\usepackage{natbib}
\usepackage[colorlinks,citecolor=blue,urlcolor=blue,bookmarks=false,hypertexnames=true]{hyperref} 
\usepackage{graphicx}

\usepackage{txfonts}
\begin{document}

   \title{From incoherent field to coherent reconnection}

  \subtitle{Understanding convection-driven coronal heating in the quiet Sun}

   \author{Rebecca A. Robinson
          \inst{1}$^,$\inst{2}
         \and
          Mats Carlsson \inst{1}$^,$\inst{2}
          \and
          Guillaume Aulanier \inst{3}$^,$\inst{1}
          }

   \institute{Rosseland Centre for Solar Physics, University of Oslo, P.O. Box 1029, Blindern, NO-0315 Oslo, Norway \\
              \and
             Institute of Theoretical Astrophysics, University of Oslo, P.O. Box 1029, Blindern, NO-0315 Oslo, Norway\\
             \and 
             Sorbonne Universit\'e, Observatoire de Paris - PSL, \'Ecole Polytechnique, Institut Polytechnique de Paris, CNRS, Laboratoire de physique des plasmas (LPP), 4 place Jussieu, F-75005 Paris, France \\
             }

   \date{}

 
  \abstract
   {  Magnetic reconnection in the quiet Sun is a phenomenon that is consistently observed, however, its conditions of occurrence are not as well known as for more energetic events. It has recently become feasible to address this issue with 3D numerical simulations of realistically stratified and convection-driven reconnection.}
   {We aim to illustrate ways by which quiet Sun fields may contribute to solar atmospheric heating via magnetic reconnection that is driven by convective motion. We also aim to compare our complex stratified model to earlier idealized coronal models in terms of reconnection drivers and topological conditions.}
   {We analyzed a simulation of the quiet Sun in which a complex coronal magnetic field is self-consistently driven by the underlying convection. We employed a selection of Lagrangian markers to trace the spatiotemporal behavior of specific magnetic features that are relevant to magnetic reconnection and atmospheric heating.}
   { A relatively large-scale reconnection-driven heating event occurs in the simulated corona, in a flattened X-shaped feature characterized by a weak field and high current. It is reminiscent of a hyperbolic flux tube (HFT), which is located at the interface between multiple flux systems. One of these is a smooth overlying horizontal field and the two most relevant others are located below the HFT. They consist of an arcade and a horizontal flux rope which eventually reconnect with the overlying field, raising coronal plasma temperatures up to 1.47 MK. }
  { We have identified a reconnection-driven coronal heating event in a quiet Sun simulation. We find that our results are in good phenomenological agreement with idealized coronal flare models, which demonstrates that the same general physical concepts are valid. However, we also find that the reconnecting flux rope and arcade are neither formed by any obvious coherent flux emergence, nor by any ordered photospheric motion or flux cancellation. Instead, they seem to develop merely from the self-consistent convective driving of pre-existing tangled field lines. This gradual and smooth ordering suggests an inverse cascade of magnetic helicity via smaller reconnection events, located at or above slowly-moving photospheric flux concentrations. We suggest that this case is representative of many heating events that may be ubiquitous in the real quiet Sun. }

   \keywords{Magnetohydrodynamics (MHD), Magnetic reconnection, Magnetic fields, Convection, Sun: atmosphere
               }

   \maketitle
%

\section{Introduction}
Magnetic reconnection is considered a ubiquitous and consistent source of solar atmospheric heating, as reconnection events have been observed and simulated under a variety of solar magnetic conditions \citep{2002mwoc.conf...47P, 2007A&A...465L..43V, 2010A&A...513A...1D}. During reconnection events in both the quiet Sun and active regions, magnetic flux makes its way from the solar interior to the photosphere, emerging into the atmosphere and dispersing energy upon connecting with a neighboring non-parallel field \citep{1957JGR....62..509P}. When this happens, stored magnetic energy is released and transformed into kinetic and thermal energy, accelerating particles and heating the local plasma (\citet{1958IAUS....6..123S, 1958NCim....8S.188S}). For a comprehensive overview of solar magnetic reconnection, we refer to the \citet{2022LRSP...19....1P} Living Review and references therein.

Several previous studies have explored the effect of flux emergence on overall magnetic topology, reconnection, and heating; whether it is in the form of a uniform flux sheet \citep{2014ApJ...788L...2A, 2014ApJ...781..126O, 2017ApJ...839...22H} or a coherent flux tube \citep{Emonet_1998, 2001ApJ...554L.111F, 2005ApJ...635.1299A, 2018ApJ...859L..26M, 2021ApJ...907...19K}. Others have shown that flux cancellation and photospheric footpoint motion are important ingredients for evolving magnetic features that can lead to reconnection \citep{1989ApJ...343..971V, 2003JGRA..108.1285P, 2005A&A...430.1067A, 2005A&A...444..961A, 2006A&A...451.1101D, 2006A&A...459..627D, 2007A&A...473..615W, 2014ApJ...787...87V} whether that prescribed motion is shearing, rotational, or converging. 

In addition to footpoint motion, several studies have quantified the energizing effects of photospheric flux loss, correlating photospheric emergence and cancellation with coronal brightening events. The flux emergence rate in the quiet Sun has been estimated using various methods and observations \citep[e.g.,][]{2011SoPh..269...13T,2013SoPh..283..273Z,2016ApJ...820...35G,2017A&A...598A..47A}. By tracking particular magnetic features in \textsc{Sunrise} observations of the photosphere, \citet{2017ApJS..229...17S} estimated the flux emergence rate in the internetwork (IN) including for the first time flux values as low as $9 \times 10^{14}$ Mx. Later, \citet{2018A&A...615L...9C} suggested that flux loss from photospheric cancellation in particular could power chromospheric reconnection and coronal heating. Then, using a coordinated observation campaign between the Swedish 1m Solar Telescope (SST) and the \textit{Interface Region Imaging Spectrograph} (IRIS), \citet{2018ApJ...857...48G} analyzed the signatures of photospheric flux cancellation in chromospheric reconnection events.

Magnetic reconnection is an effective way to order magnetic fields and reallocate the energy budget of solar plasma but it is not always clear what the specific driving mechanisms behind this reconnection are. Simple topological experiments have demonstrated how footpoint motion can result in rising magnetic arcs that eventually reconnect with a non-parallel overlying field \citep{2003ApJ...595..506G, 2005A&A...444..961A, 2006SoPh..238..347A, 2016SoPh..291..143E}, but with more complicated topologies, it is challenging to isolate the exact mechanics of particular footpoint motion in relationship to flux emergence and possible reconnection. 

Furthermore, the ways by which magnetic fields order themselves under reconnection depends on a combination of the overall topology, the angle of reconnection, and the footpoint driving. \citet{2015ApJ...805...61Z} have shown in simplified plane-parallel loop simulations that co-rotating footpoints result in magnetic reconnection and ordering as an ``inverse cascade'' of helicity, meaning the helicity of the field is transferred to larger scales rather than smaller scales. Similar studies have tested and verified this phenomenon \citep{2017ApJ...835...85K, 2019ApJ...883..148R}, but this has yet to be developed and analyzed in the context of more complicated magnetic features such as those generated by convection. 
In our study, we have no flux sheet insertion nor coherent flux emergence, nor do we prescribe explicit footpoint motion. Instead, our simulated reconnection event is self-consistently driven by motion in the simulated convection zone. Previously, \citet{2008ApJ...679L..57I} noted the impact of convection on otherwise uniform magnetic fields and \citet{Li_2022} recently reported on convection-driven reconnection events in their simulations of microflares. Convection-driven heating has also been seen before in earlier \textit{Bifrost} simulations, so described in \citet{2015ApJ...811..106H}, who explored various characteristics of large- and small-scale heating. 

Our simulated reconnection is characterized by two arched magnetic loops, one arcade and one flux rope, both interacting with an overlying horizontal field. Such features are evidenced by coronal current sheets, which indicate the presence of null-less quasi-separatrix layers (QSLs), as introduced by \citet{1996A&A...308..643D} and explored further in \citet{2005A&A...444..961A} and \citet{2006SoPh..238..347A}. Both magnetic features are firmly rooted in the photosphere on one side, and these photospheric roots do not undergo complete flux cancellation during the reconnection process. 

In this paper, we describe the magnetic characteristics and evolution of this convection-driven event, drawing the conclusion that neither coherent flux emergence nor specific footpoint motion are necessary for magnetic reconnection in the quiet Sun. Instead, we explore the possibility that the build-up of the pre-reconnection loops is a result, to some degree, of the inverse helicity cascade through major and minor reconnection events, although further evidence is required in a future study. We suggest that the processes behind ordering pre-existing coronal lines may contribute to the build-up of strong reconnection events, in addition to more traditional magnetoconvective processes.

This paper is primarily focused on the magnetic geometry of a particularly strong reconnection event, and present this case as the first in a series of studies which stem from the same simulation. The specifics of inverse helicity cascade as well as observational signatures of our reconnection event will be discussed in follow-up papers.


\section{Methods}\label{sim}

\subsection{The \textit{Bifrost} code}
The parallel numerical code \textit{Bifrost} explicitly solves the MHD partial differential equations, and it is described in detail in \citet{2011A&A...531A.154G}. The code employs a sixth order differential operator, a fifth order interpolator, and diffusive terms. The architecture of the code is designed for optimal user-friendliness, employing a single input file which allows for simple parameter modifications. Users can also select several different modules; for example, test particles, equations of state, radiative transfer, and boundary conditions can all be easily modified via one comprehensive input file. Time stepping in our simulation follows the explicit, third-order recipe detailed in \citet{hyman}.

Our \textit{Bifrost} simulation box includes a slice of convective solar interior that extends from the bottom boundary to the $\tau_{500} = 1$ surface. We note that we define the height z = 0 Mm as the average solar surface; specifically, the mean height of the $\tau_{500} = 1$ surface. In \textit{Bifrost} coordinates, the vertical component is depth rather than height such that it increases from the coronal boundary down to the convective boundary. All vector quantities are aligned with this geometry. We note that our description of height in the text is given with respect to the average solar surface, although the vertical component in our 3D renderings is aligned with the \textit{Bifrost} geometry. 

Radiative transfer in the photosphere and lower chromosphere use multi-group opacities with four opacity bins \citep{1982A&A...107....1N}, including scattering \citep{2000ApJ...536..465S}, and is implemented using a short characteristics scheme following \citet{2010A&A...517A..49H}. Hydrogen is treated in local thermodynamic equilibrium (LTE) for our experiment. The overlying atmosphere includes a dynamic upper chromosphere, transition region, and corona; the radiative energy budget of which is described in detail in \citet{mats}. Thermal conduction along coronal magnetic field lines is integral to the energy budget of the upper atmosphere, calculated as detailed for the MURaM code in \citet{Rempel_2016}.

The diffusive term in \textit{Bifrost} is split into two parts: one global term and one localized (hyper-diffusion) term, which is larger for local gridpoints where artificial diffusion proves necessary \citep{2011A&A...531A.154G}. The larger hyper-diffusion term prevents gradients that are too high to handle numerically; for example, the dissipation of current sheets to scales that are smaller than the gridpoint size. Diffusive parameters such as the Reynolds or Magnetic Reynolds number then have a range of values in space and time instead of just one global value, and features that form where gradients are highest are artificially diffused such that their geometries do not breach the simulation resolution. We note that the hyper-diffusion term regulates the ways by which magnetic reconnection is possible in the simulation, governing current sheet formation and reconnection processes. 

In this study, we run a quiet Sun simulation with a horizontal extent of 12 Mm and a horizontal pixel resolution of 23 km. The simulation is a continuation of a simulation described in \citet{2019ApJ...878...40M}, which had a horizontal extent of 6 Mm and a horizontal pixel resolution of 5 km. This reference simulation was first degraded to a resolution of 10 km and then to 20 km, and we define our t = 0 at the start of the 20 km horizontal resolution simulation. At t = 240 s, this simulation box was duplicated in both horizontal directions and resampled to a 512$^3$ grid and a horizontal pixel resolution of 23 km. The vertical extent of the computational box ranges from 2.5 Mm beneath the average solar surface to 8 Mm above the average solar surface, with nonuniform vertical spacing ranging from 30 km at the bottom boundary in the convection zone to a finer 12-14 km between the average solar surface and 4 Mm above it, then increasing to a coarser 70.5 km at the coronal boundary.  

In \textit{Bifrost}, the magnetic field is initialized as a free parameter. The original simulation had a magnetic field resulting from a small-scale dynamo starting from a uniform vertical field of 2.5 G. To simulate a quiet atmosphere in our simulation, the resampled box was reinitialized with a potential field extrapolation from a balanced vertical magnetic field at the bottom boundary, with a similar amplitude to that of the original box.  

In this experiment, boundary conditions are horizontally periodic because this allows for a well-behaved simulation box and does not cause reflections at the horizontal boundaries. We note that the consequences of these boundary conditions include magnetic connectivity across the boundaries as well as the formation of functionally static horizontal fields in the corona, both of which are relevant to the analysis of the simulation we present in this paper. The lower boundary is open to flows with a prescribed entropy such that any inflowing material maintains an effective temperature of roughly 5780~K. The upper boundary is kept open.

\subsection{Visualizing the magnetic field}\label{vaporsec}
Since convection alone drives the evolution of the field, photospheric footpoints are always moving and the scale height of their corresponding loops is constantly changing. With that, we expect reconnection events to occur throughout the duration of the simulation simply due to the statistical likelihood of non-parallel field interactions. A detailed analysis of one particular reconnection event will be given in the following sections, but a description of the methods we used to trace and follow reconnecting field lines is relevant here.

Using the Visualization and Analysis Platform for Ocean, Atmosphere, and Solar Researchers (VAPOR) software \citep{vapor}, we traced the flow of the magnetic field as it evolves throughout the simulation. In VAPOR, users have several options regarding how to seed a flow field for tracing. For example, users may choose a random distribution of seeds throughout the entirety of the simulation cube, a uniform grid of seeds, pre-selected seeds from a file of coordinates, or a random distribution of seeds biased toward either high or low values of a selected variable. In addition, users may specify a smaller region within the cube where the field should be seeded, ignoring all other gridpoints in the cube.

The resulting field tracing is then a direct function of the seed locations, so it is necessary for the user to choose a seeding method that selects the most relevant fields to the research question. In our case, we chose to bias the field tracer toward certain variables in order to pick out certain components of the magnetic field. In VAPOR, biasing a random distribution of seeds means that the software is more likely to select a certain gridpoint as a seed if the gridpoint meets the criteria of the bias. With a positive bias, the seed distribution favors high values of the selected variable. With a negative bias, the seed distribution favors low values of the selected variable \citep{vapor}.

This seeding method is sufficient for looking at still pictures within our simulation, but it is not sufficient to follow the time evolution of particular field lines because the output cadence of the simulation is ten solar seconds. As our simulation timestep is on the order of $10^{-3}$ s, each output file then represents the end result of several thousand simulation updates. This cadence is not sufficient to follow the field evolution over time between output files because the field evolves considerably over thousands of simulation updates. Therefore, using the seeding methods in VAPOR to trace the field (even with a consistent bias) is not sufficient for following specific magnetic field lines over time. The output cadence is also too infrequent to rely on advecting the magnetic field with the velocity field to update the field lines. To address the problem of constructing a reliable time series of magnetic field evolution, we employ a test particle module in \textit{Bifrost}.

\subsection{The \texttt{corks} module}
The ability to utilize optional modules makes \textit{Bifrost} a versatile and useful code for solving specific problems. In our case, adding Lagrangian markers to our simulation simplified our analysis of the dynamic magnetic features that develop due to convection.

To find out how particular magnetic field lines evolve in the simulation, we employed a test particle \textit{Bifrost} module called \texttt{corks} \citep{2018A&A...614A.110Z, 2022A&A...665A...6D}. The \texttt{corks} module was designed as a way to introduce Lagrangian markers to \textit{Bifrost} simulations without affecting the physics of the simulation, and it eliminates the need to increase our output cadence in order to adequately advect the magnetic field with the velocity field to trace magnetic field lines in time. Instead, we can simply follow the ``corks'' as they follow the flow of the simulation and update their locations with each simulation timestep. We initiated one cork per gridpoint at t = 9\,669 s with no additional cork injection, meaning the number of corks in the box (512$^3$) never changes throughout the rest of the simulation. 

The \texttt{corks} module enabled us to trace individual markers and visualize the magnetic field lines associated with them at any given simulation timestamp. By selecting corks located within specific magnetic features and using them as seeds, we traced corresponding field lines in order to characterize those magnetic features as they evolved in time. This means that we used the cork coordinates for each output file as a set of pre-selected seeds, which we then loaded into VAPOR for tracing. Furthermore, corks that were spatially associated with current sheets or reconnecting field lines served as seeds for tracing any magnetic fields relevant to their corresponding reconnection region. No corks are directly associated with reconnection regions because the \texttt{corks} module assumes that the field is frozen into the surrounding plasma, but corks that are co-located with reconnecting field lines are still useful. We refer to Section \ref{csel2} for an extended description of our cork selections as they relate to the simulation results. 

Once we isolated specific topological features by following the motions of selected corks and tracing the corresponding field lines, we had a clearer picture of how the magnetic field moves, orders itself, reconnects, and develops throughout this simulation. An interpretation of the field evolution and its implications is given in the following sections.

\section{Results}\label{res}
In order to understand quiet Sun field strengths and distributions as a stepping stone to understanding atmospheric heating, we designed a simulation to imitate a featureless quiet Sun atmosphere. In this case, ``featureless'' does not imply a blank or uniform photosphere, but rather one that is shaped by convection alone. The results of that simulation formed the basis for our cork selection which informed our understanding of the dynamic magnetic features that emerged during the simulation, as well as the subsequent reconnection and heating.

\begin{figure*}
    \centering
    \includegraphics[width=0.8\linewidth]{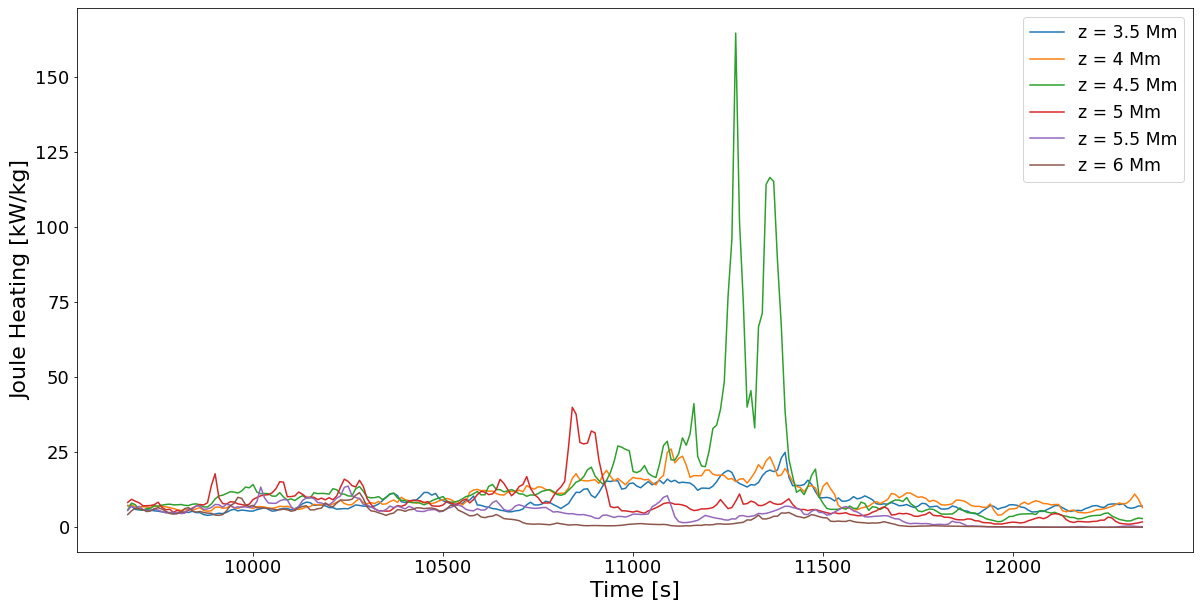}
\caption{Time series of average Joule heating at multiple heights. Several impulsive heating events are clearly seen here, especially at a height of 4.5 Mm (green line).}
\label{fig:qtavg_cx}
\end{figure*}

\begin{figure*}
\begin{subfigure}{.49\textwidth}
  \centering
  \includegraphics[width=\linewidth]{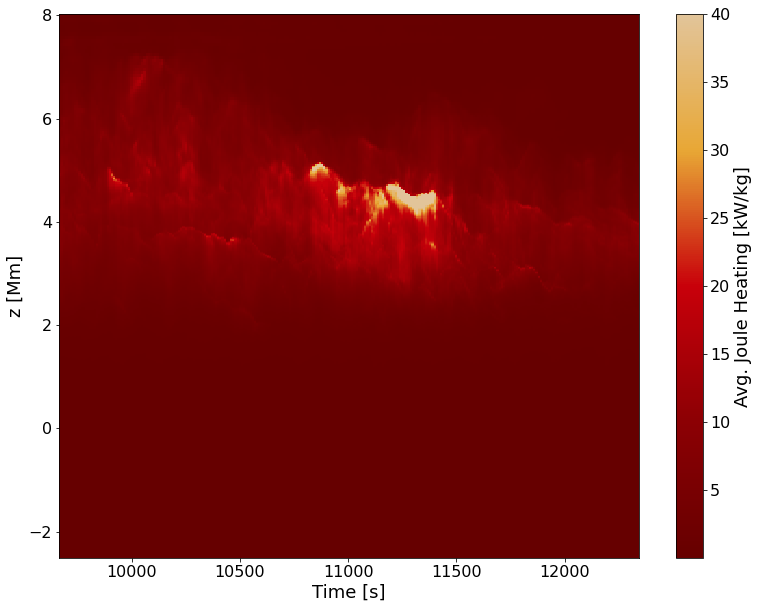}  
  \label{fig:heattime}
\end{subfigure}
\begin{subfigure}{.49\textwidth}
  \centering
  \includegraphics[width=\linewidth]{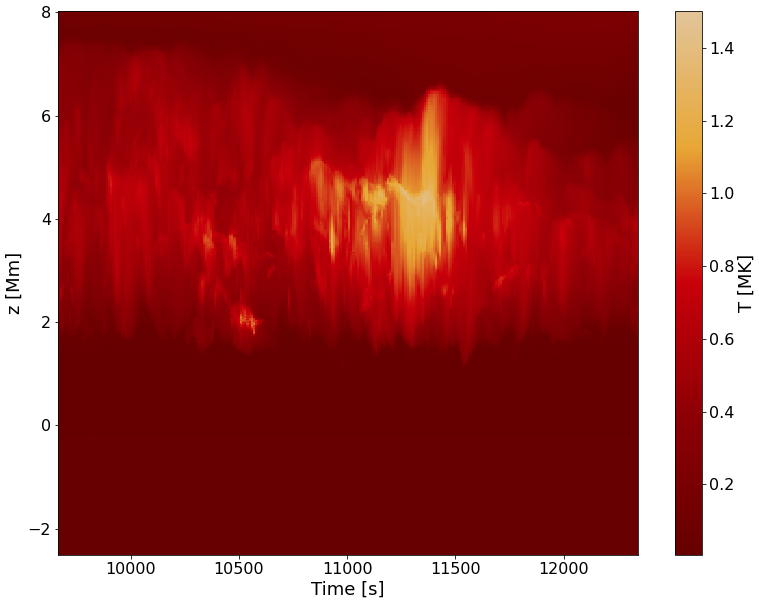}  
  \label{fig:temptime}
\end{subfigure}
\caption{Time series of horizontally averaged Joule heating (kW/kg) for all heights (left). Time series of maximum horizontal temperature (MK) for all heights (right). }
\label{fig:heattemp}
\end{figure*}

\subsection{Atmospheric heating}
\textit{Bifrost} includes calculations of Joule heating which (physically) is a result of energy dissipation of current sheets that develop where there are strong gradients in the magnetic field. With that, strong Joule heating events act as pointers for the locations of dissipating current sheets in time and space, which themselves act as proxies for the locations of reconnecting field lines. Localized heating bursts in the simulation are therefore strong indicators of reconnection regions.

During the course of the run we see several localized heating events in the simulated atmosphere. Figure \ref{fig:qtavg_cx} illustrates the time-varying horizontally-averaged Joule heating at several different heights. We note that the units of Joule heating are kW/kg, compensating for the very diffuse nature of the corona and emphasizing coronal heating with energy per particle rather than energy per volume. It is clear from Figure \ref{fig:qtavg_cx} that several impulsive heating events occur at multiple heights throughout the simulation, as evidenced by the many peaks leading to the strongest events. Of the selected heights in Figure \ref{fig:qtavg_cx}, it is clear that the most powerful heating events occur at a height of 4.5 Mm, between roughly t = 11\,200 s and t = 11\,500 s. To see the heating events more clearly for all heights, Figure \ref{fig:heattemp} shows a time series of horizontally averaged Joule heating (left panel) and maximum horizontal temperature (right panel) as a function of height. In the left panel of Figure \ref{fig:heattemp}, we see evidence of the same impulsive heating events as in Figure \ref{fig:qtavg_cx}, indicating that the simulated atmosphere is heated locally in several places throughout the run. Most notably, there is an obvious increase in Joule heating per particle between t = 10\,800 s and t = 11\,500 s, at heights between 3.5 Mm and 5.5 Mm above the average solar surface. Corresponding to those times and locations, there is a jump in atmospheric temperature up to 1.47 MK, as shown in the right panel of Figure \ref{fig:heattemp}. During the time period between t = 11\,200 s and t = 11\,500 s (corresponding to the two largest peaks in Figure \ref{fig:qtavg_cx}) and within the volume enclosed by the entire horizontal extent and heights between 4.5 Mm and 6 Mm, Joule heating alone accounts for an energy of roughly $5.4 \times 10^{17}$ J. This is comparable to typical nanoflare energies of $\approx 10^{17}$ J, placing this event closer to the nanoflare regime than the microflare regime \citep{2011SSRv..159..263H}. These figures are useful for localizing heating events in time and space, but exploring a 3D spatial rendering of the heating at one specific time would shed some light on its source. 

\begin{figure*}
\begin{subfigure}{.49\textwidth}
  \centering
  \includegraphics[width=\linewidth]{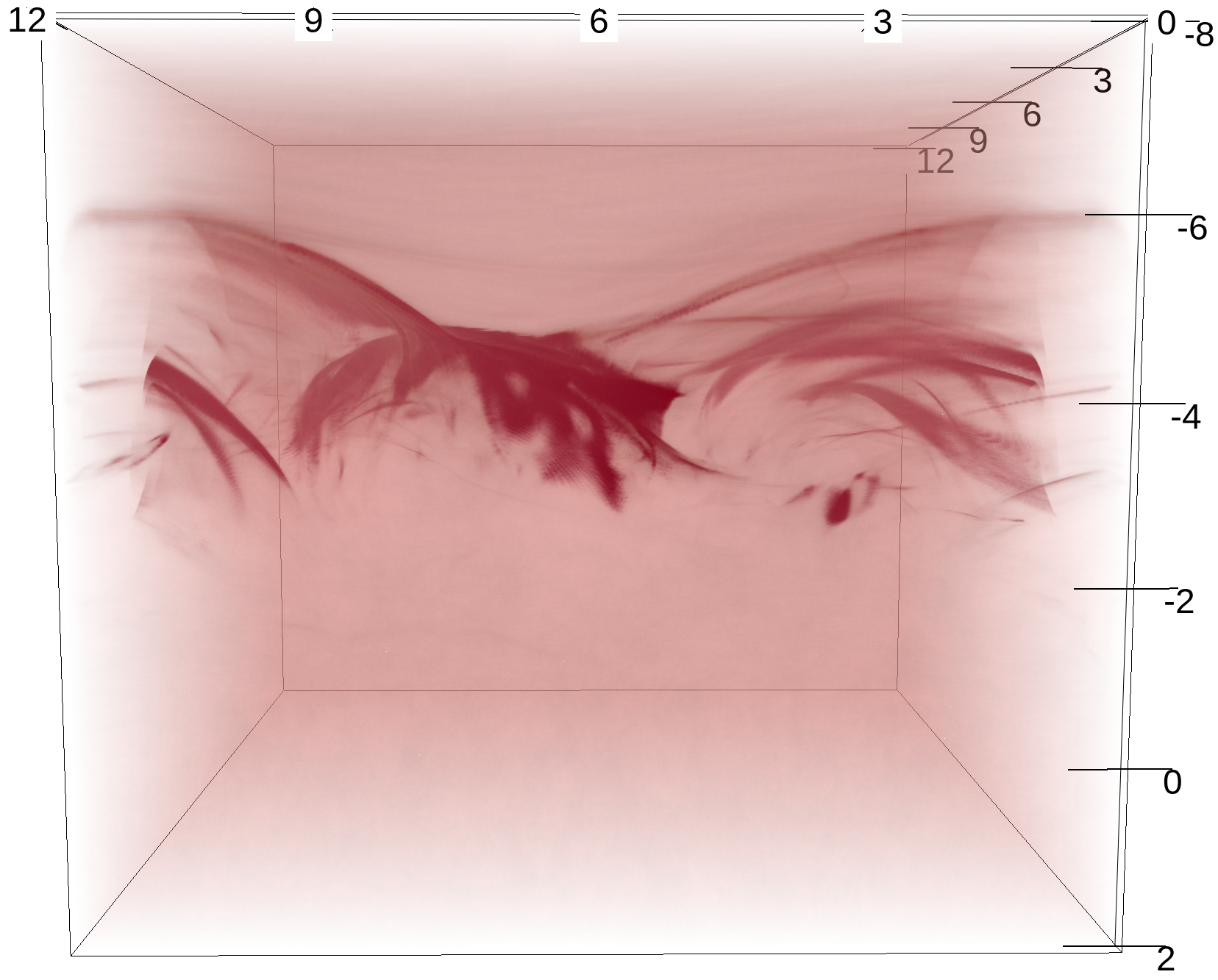}  
  \label{fig:sheet_cx}
\end{subfigure}
\begin{subfigure}{.49\textwidth}
  \centering
  \includegraphics[width=\linewidth]{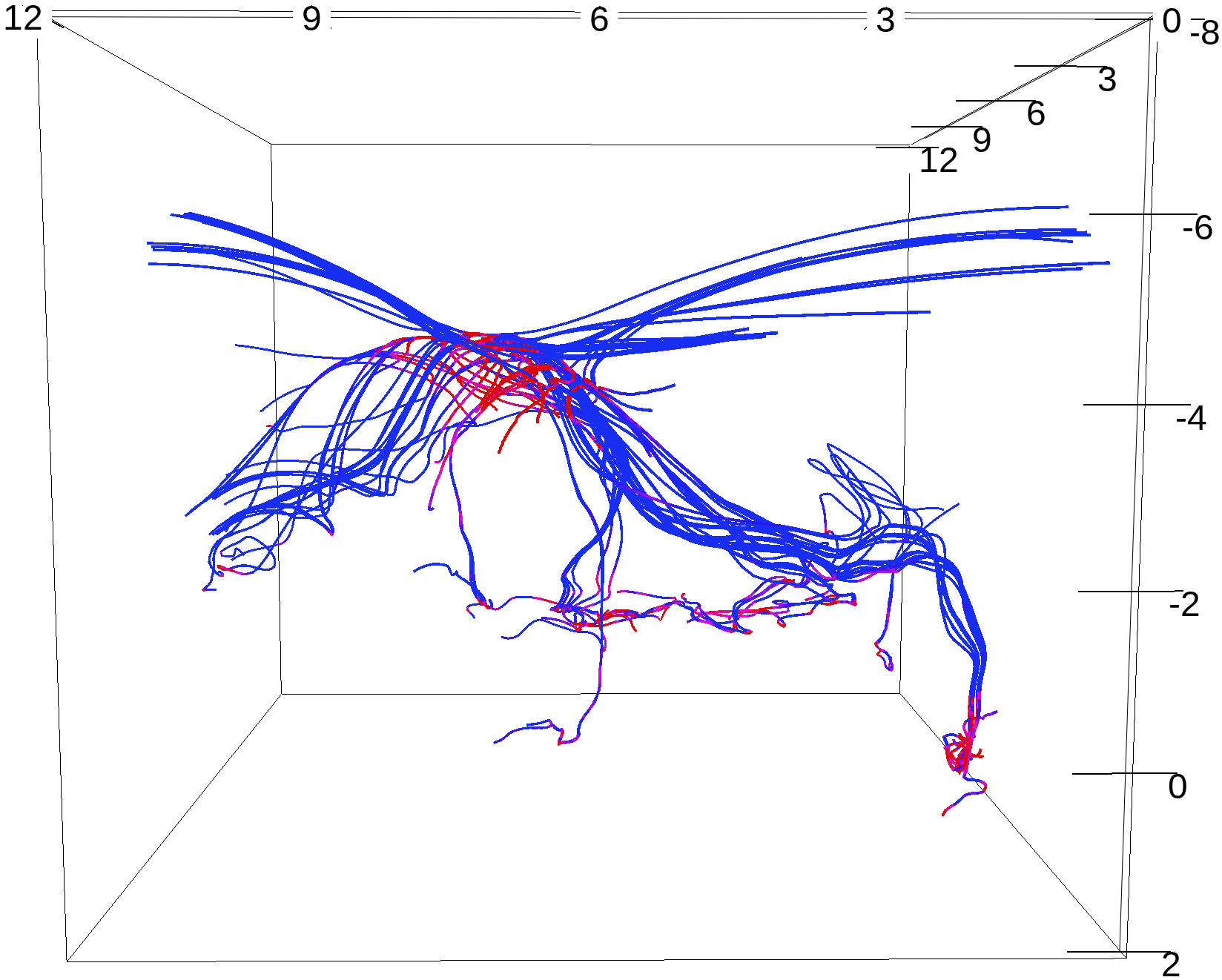}  
  \label{fig:job}
\end{subfigure}
\caption{Volume rendering of Joule heating at t = 11\,360 s (left). Magnetic field tracing at t = 11\,360 s with a strong positive bias towards $\vec{\jmath}/\vec{B}$ at coronal heights above 3.5 Mm (right). The traced lines are colored by $\vec{\jmath}/\vec{B}$, red at highest values and blue at lowest values. There are high values of $\vec{\jmath}/\vec{B}$ at the reconnection region and also deep in the convection zone (see flux concentration on the right side of the box) but the tracing is seeded only by coronal seeds with high $\vec{\jmath}/\vec{B}$ values.}
\label{fig:sheet_job}
\end{figure*}

The left panel of Figure \ref{fig:sheet_job} shows a well-localized dissipating current sheet in the corona at t=11\,360 s. This is a volume rendering of Joule heating, such that the red features highlight areas where current is being dissipated as heat in the simulation. This panel represents one selected column of the left panel of Figure \ref{fig:heattemp} at t=11\,360 s, expanded into three spatial dimensions to illustrate the shape of the dissipating current sheet. The X-shaped geometry of this current sheet is notable as it consists of a central spine and then several horizontal jet-like structures that seem to emanate from it, similar to current sheets near separatrices as described in \citet{BULANOV1995219}. Here, it is useful to recall that the magnetic field prescribes the current in this simulation, meaning this dissipation is a result of some rapid change in the coronal magnetic field. 

By understanding this structure and in looking at the geometry of the heating, we determined that this type of current sheet dissipation was likely associated with a magnetic reconnection event. To understand the reconnection in terms of magnetic field shape and evolution, we selected specific corks and used them as seeds to trace the magnetic field in time and space. 

\subsection{Magnetic topology and reconnection}
To understand the magnetic features that are relevant to such rapid changes in heat dissipation and temperature, it is important to select an optimal set of corks that are nearest the lines associated with reconnection. Doing so requires an understanding of the conditions for reconnection as well as the relevant parameters such as current density and field strength. The cork selection process and the associated field tracings are discussed in the following subsections. 

\subsubsection{Physics-informed cork selection}\label{csel2}
Considering that the box is full of changing magnetic fields driven by convective motion, reconnection events are expected to happen frequently and the most straightforward way to find them is to look for impulsive heating events, as discussed above. Based on Figure \ref{fig:heattemp}, the most energetic heating events occur in the corona between t=10\,800 s and t=11\,500 s. The times and locations of enhanced heating provide a baseline for selecting relevant corks which correspond to dissipating current and possible reconnection regions. For the duration of the analysis, we used t=11\,360 s as a reference timestamp as in Figure \ref{fig:sheet_job}, which we set as the basis for our cork selection. 

As per the X-shaped current sheet dissipation shown in the left panel of Figure \ref{fig:sheet_job}, there must be some rapid changes in the coronal magnetic field at that time that would generate a current sheet and dissipate heat. This X-shaped geometry is indicative of thin current sheet formation near a hyperbolic null line in a reconnection region \citep{1966SvA....10..270S, imshennik1967two, BULANOV1995219, bezrodnykh}. With our 3D treatment, we define our reconnection region with quasi-separatrix layers (QSLs) rather than separatrices because where null lines are absent, current formation occurs along quasi-separators (aka. hyperbolic flux tubes (HFTs)). Here, the X-shaped geometry in 3D points to thin current sheet formation along a HFT. 

Picking out the largest changes in magnetic field and subsequent current sheet formation near the minimum surface is most straightforwardly done by calculating the ratio of current density to magnetic field strength, $\vec{\jmath}/\vec{B}$. A combination of small field values along the minimum surface and high current density in current-forming regions maximizes this parameter, which selects for possible reconnection regions. 

By calculating $\vec{\jmath}/\vec{B}$, we could pick out the thinnest, most significant current sheets near a reconnecting surface and see their geometries directly. As mentioned in Section \ref{vaporsec}, it is possible to bias a magnetic field tracer toward a certain variable. By choosing a strong positive bias towards $\vec{\jmath}/\vec{B}$ and focusing the bias on heights greater than 3.5 Mm, we can clearly see the geometries of reconnecting field lines in that region. A field drawn at t = 11\,360 s with the aforementioned bias is shown in the right panel of Figure \ref{fig:sheet_job}. We recall that this is not something that we would have been able to see simply by following random corks because the \texttt{corks} module assumes that the magnetic field is smoothly advected by the velocity field and does not account for possible magnetic reconnection events. Therefore, it becomes necessary to seed the field with a bias towards $\vec{\jmath}/\vec{B}$ and then select appropriate corks based on the corresponding lines, as shown in the right panel of Figure \ref{fig:sheet_job}. By finding out which magnetic field lines correspond to the highest values of $\vec{\jmath}/\vec{B}$ in the corona at that time, we selected the nearest corks to those field lines and used them as corresponding seeds. 

Since Figure \ref{fig:sheet_job} illustrates heating associated with current sheet dissipation as well as magnetic reconnection at that time, calculating the current density $\vec{\jmath}$ directly is helpful for determining where exactly the current sheets are being generated in the box. The top-left panel of Figure \ref{fig:jz} shows a horizontal slice through the box at 4 Mm above the solar surface, colored by $\jmath_z$, which is enhanced as a result of strong gradients in the horizontal magnetic field. As denoted in the figure, we have two ovular current sheet configurations which implies that there are two intersecting QSLs and therefore two associated magnetic features which intersect there. With that, corks in the vicinity of those two ovals can be used as seeds for tracing those two magnetic features.

By combining the corks selected based on their vicinity to high values of $\jmath_z$ as well as lines corresponding to high values of $\vec{\jmath}/\vec{B}$, we could follow the evolution of certain field lines that are relevant to current formation as well as reconnection. Cork selection was necessary to follow the past and future of the lines, because simply using the same biases for each separate timestep offers no temporal integrity. By using the corks, we could construct a timeline of field evolution before, during, and after reconnection. 

\subsubsection{Topological features}
\begin{figure*}
\begin{subfigure}{.49\textwidth}
  \centering
  \includegraphics[width=\linewidth]{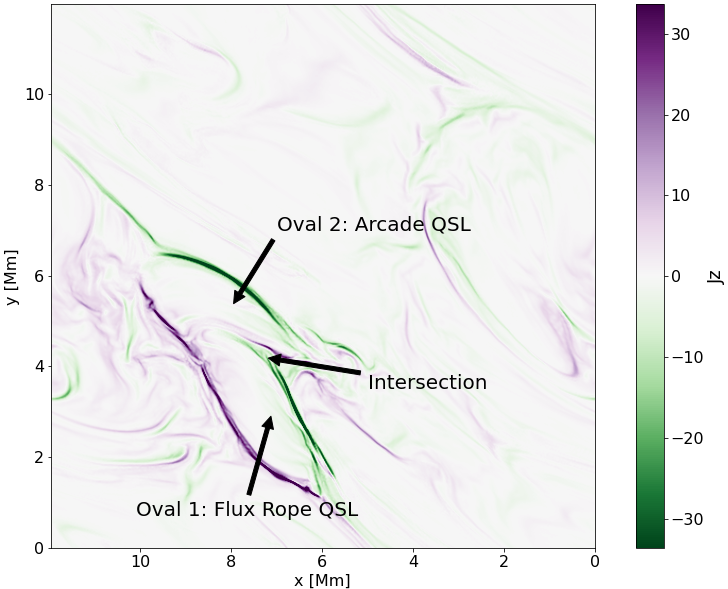}
  \label{fig:jzog}
\end{subfigure}
\begin{subfigure}{.49\textwidth}
  \centering
  \includegraphics[width=\linewidth]{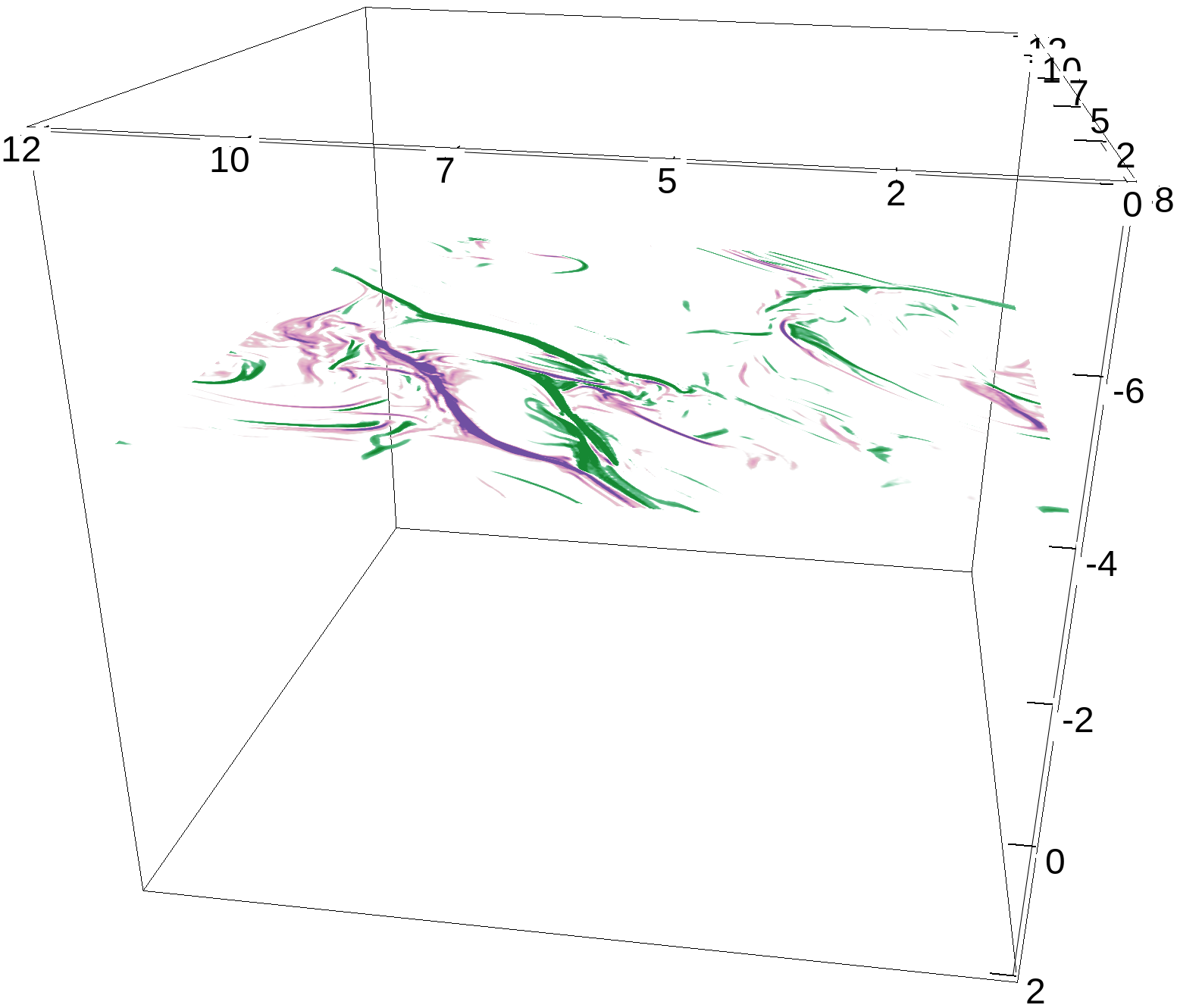}  
  \label{fig:jz3d}
\end{subfigure}
\begin{subfigure}{.49\textwidth}
  \centering
  \includegraphics[width=\linewidth]{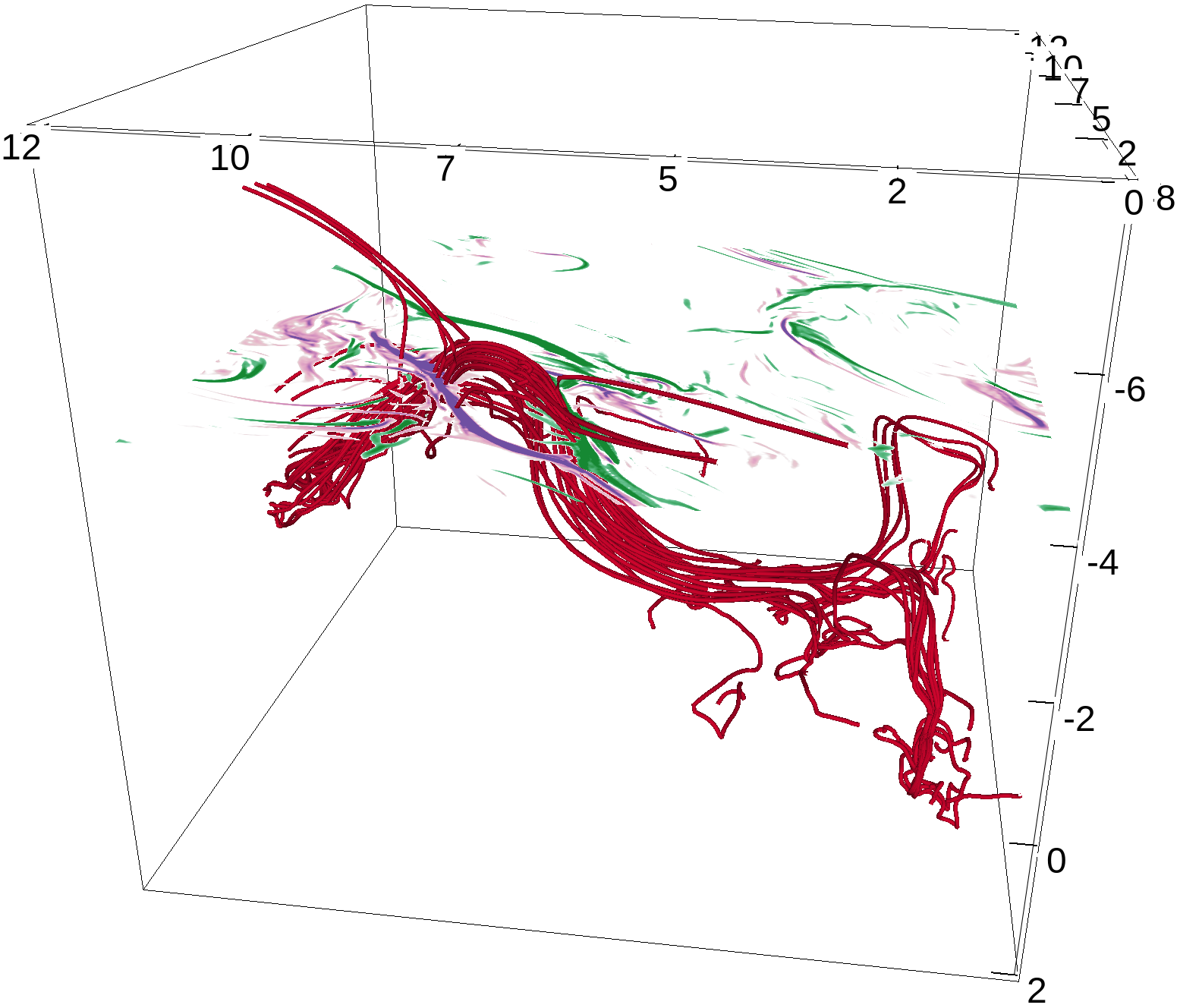}
  \label{fig:jzrope}
\end{subfigure}
\begin{subfigure}{.49\textwidth}
  \centering
  \includegraphics[width=\linewidth]{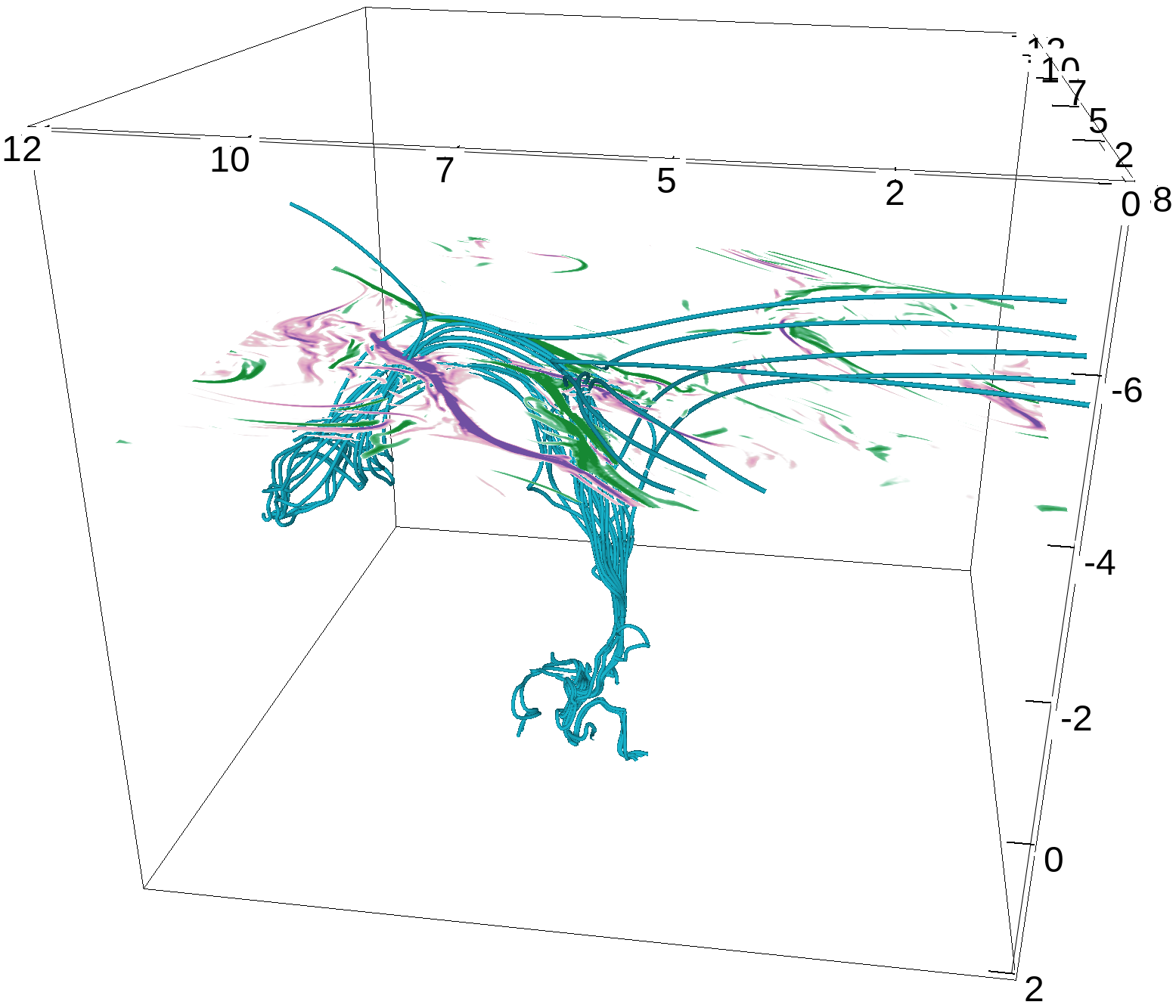}  
  \label{fig:jzarcade}
\end{subfigure}
\caption{Horizontal slice through the simulation showing $\jmath_z$ at t = 11\,360 s and at a height of 4 Mm in 2D (top-left) and 3D (top-right and bottom panels). The component $\jmath_z$ shows multiple QSLs that are associated with the magnetic features responsible for reconnection. The lower oval-shaped QSL is associated with a horizontal flux rope, as denoted with an arrow on the top-left panel and illustrated with red lines in the bottom-left panel. The upper oval-shaped QSL is associated with a magnetic arcade as denoted with an arrow in the top-left panel and illustrated with cyan lines in the bottom-right panel. The intersection between these features is located at the intersection between the two QSLs, as denoted with an arrow in the top-left panel. }
\label{fig:jz}
\end{figure*}

The selected corks discussed above provide seeds for magnetic field lines that are associated with current sheets and magnetic reconnection. As shown in Figure \ref{fig:jz}, the two ovals of current indicate two distinguishable magnetic features, and by using the selected corks as seeds, it becomes possible to determine the characteristics of those features.

The bottom two panels of Figure \ref{fig:jz} show the two distinguishable magnetic features as they relate to $\jmath_z$. The first feature is a horizontal flux rope, denoted with red lines and corresponding to the lower oval-shaped QSL as also seen in the top-left panel of Figure \ref{fig:jz}. The second feature is a magnetic arcade, denoted with cyan lines and corresponding to the upper oval-shaped QSL as also seen in the top-left panel of Figure \ref{fig:jz}. These two magnetic features are not only implied by the existence of two intersecting QSLs, but they also correspond to regions with high coronal $\vec{\jmath}/\vec{B}$ values. For the duration of the analysis, we consider these two features to be the features most illustrative of the main reconnection and heating event in the simulation. 

\begin{figure*}
\begin{subfigure}{.5\textwidth}
  \centering
  \includegraphics[width=\linewidth]{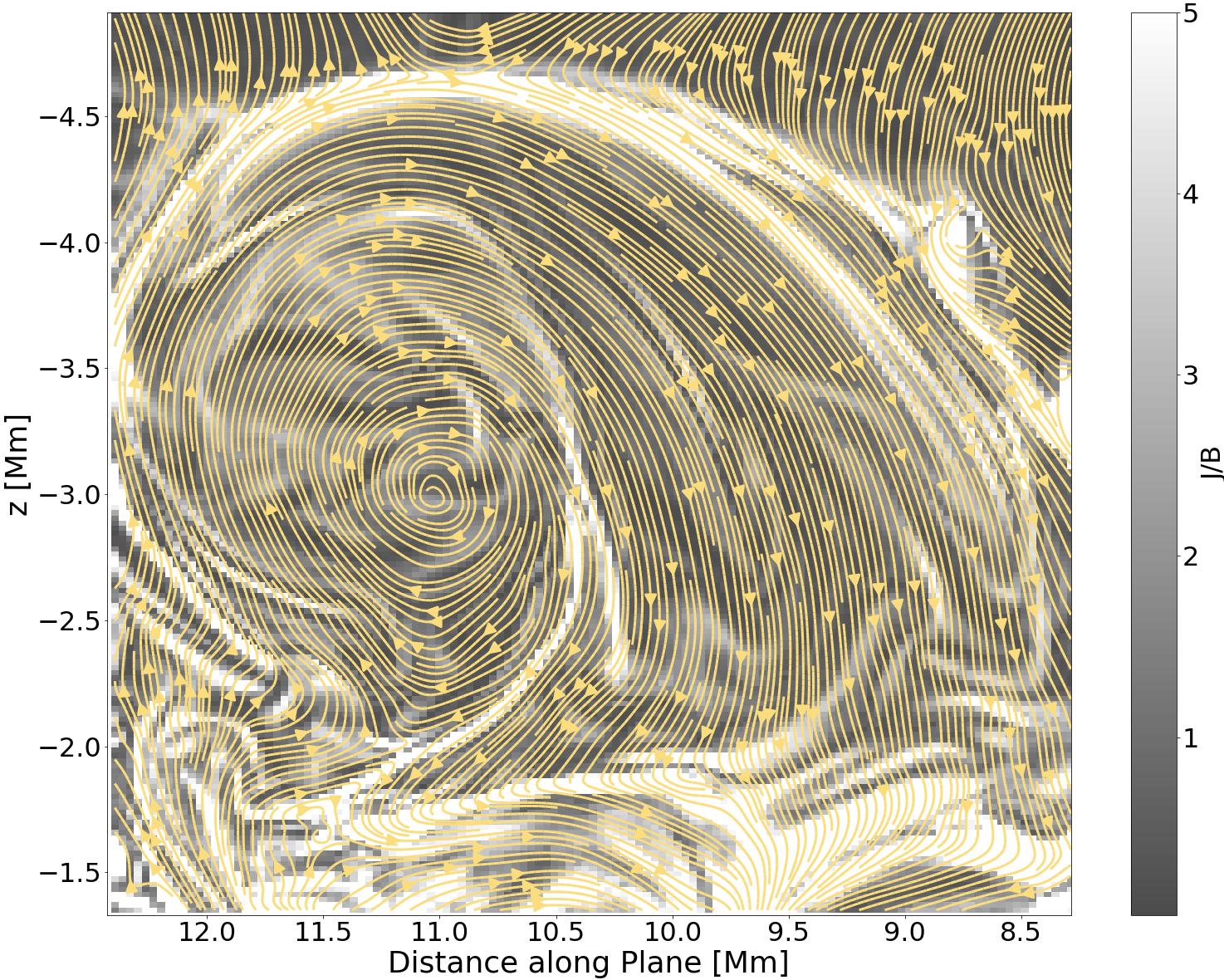}  
  \label{fig:streamplot}
\end{subfigure}
\begin{subfigure}{.52\textwidth}
  \centering
  \includegraphics[width=\linewidth]{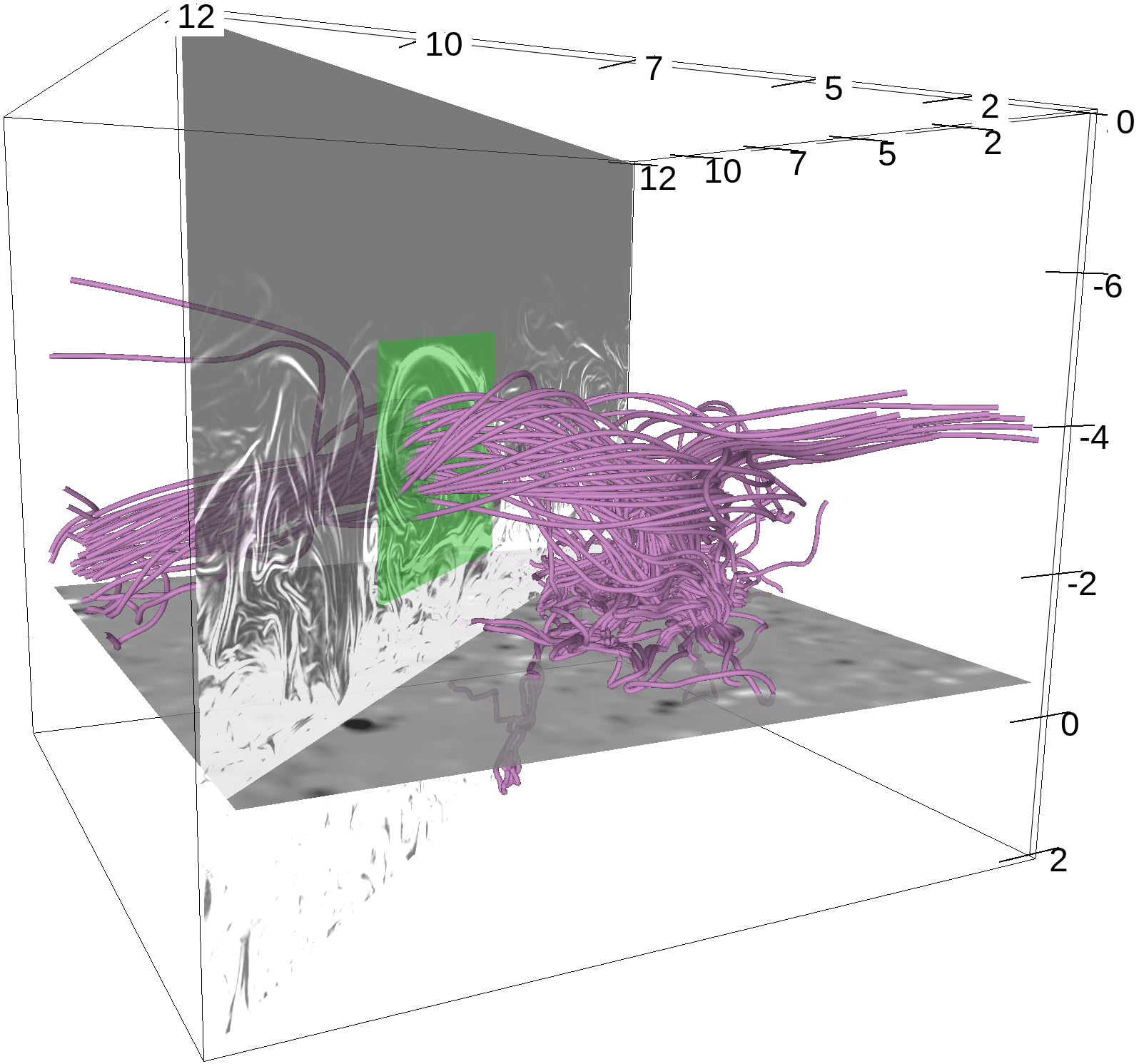} 
  \label{fig:fullflux}
\end{subfigure}
\caption{Orthogonal slice through the flux rope (shown in the bottom-left panel of Figure \ref{fig:jz}) to illustrate the twist of the field (yellow arrows) as well as $\vec{\jmath}/\vec{B}$ values in the background at t = 11\,240 s, before the main reconnection event (left). Full extent of the simulation box including a more comprehensive picture of the flux rope, with the orthogonal slice of $\vec{\jmath}/\vec{B}$ values to indicate where the cut was taken (right). The green rectangle indicates the field of view of the left panel.}
\label{fig:stream}
\end{figure*}

To illustrate that the feature shown in the bottom-left panel of Figure \ref{fig:jz} is indeed a part of a flux rope, we took an orthogonal slice through those lines at t = 11\,240 s, a time when the rope has been formed but the reconnection event has not yet occurred. The left panel of Figure \ref{fig:stream} shows the magnetic field vector along this orthogonal plane, with $\vec{\jmath}/\vec{B}$ values shown in the background. Not only are there multiple thin current sheets inside the rope, but the scale of the overall twist at this time is obvious. This twist illustrates that the extent of the flux rope is far larger than the representative lines traced by markers, but the feature is indeed part of a larger flux rope. The right panel of Figure \ref{fig:stream} shows a 3D rendering of the full extent of the flux rope passing through the slice, along with the $\vec{\jmath}/\vec{B}$ plane given for reference.

While the aforementioned features are relevant to the reconnecting region, they are not the only features that contribute to the overall topology of the box. Those two features are also clearly in the process of reconnecting with another field which is not shown by simply tracing the selected corks, but evidenced by the kinked geometry of some of the lines. The lines that are traced depend entirely on the selected seeds; thus, in order to see more features and get a more comprehensive view of the field, it is necessary to seed the field with varying biases.  

\begin{figure*}
    \centering
    \includegraphics[width=\hsize]{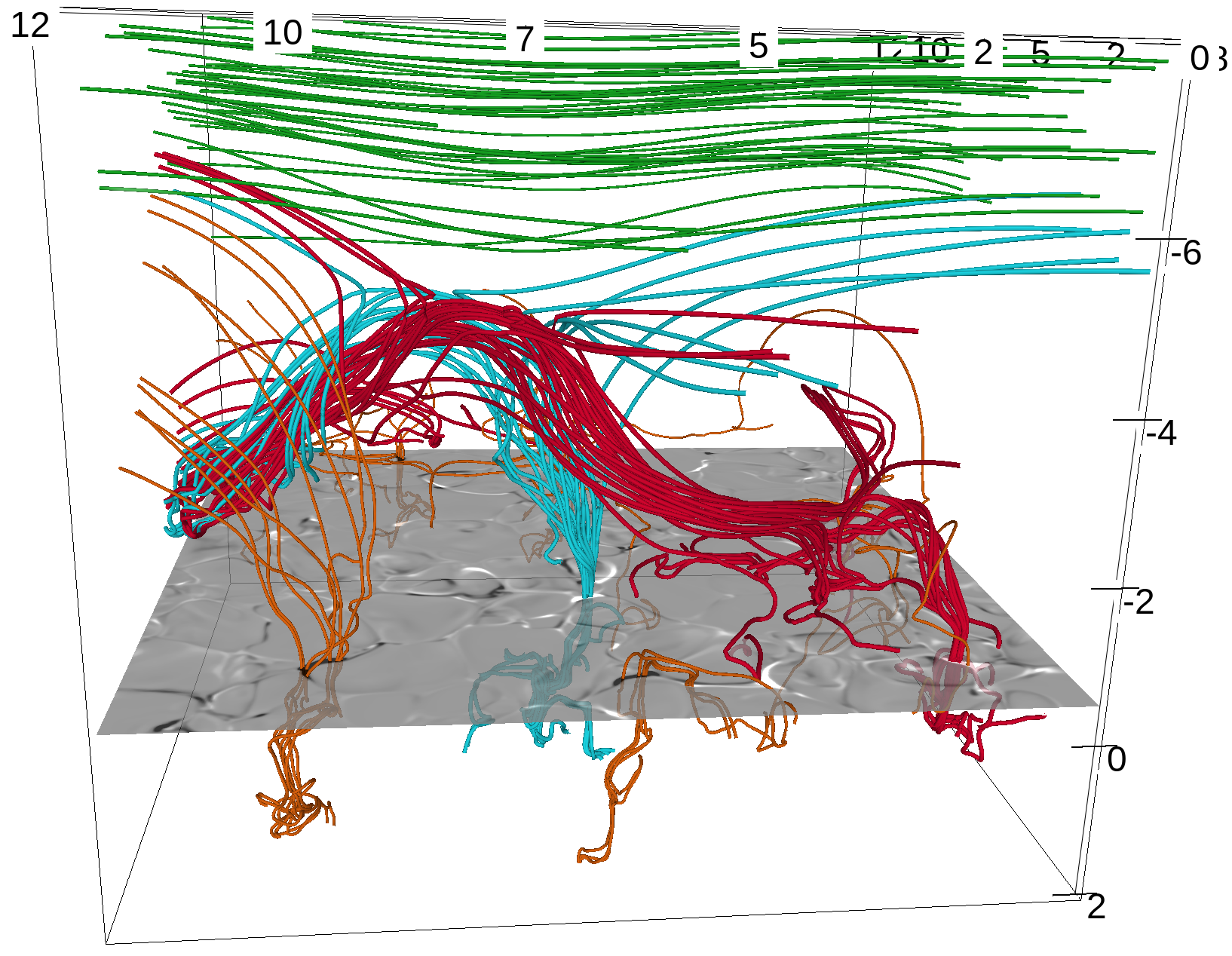}
    \caption{Visualization of field lines at t = 11\,360 s with the photospheric B$_z$ field and the overlying magnetic field seeded four different ways. The green field represents the overlying horizontal coronal field, the red and cyan fields are seeded by corks near regions with high current density and $\vec{\jmath}/\vec{B}$, and the orange field is seeded with a bias toward the photospheric $-B_z$ field. The overlying horizontal field (green) is nearly anti-parallel to the red and cyan features.}
    \label{fig:bfield_comps}
\end{figure*}

To find additional field components besides the two features traced by corks, we looked above and below the reconnection region for other fields with which the two features reconnect as well as photospheric connections that are not associated with the two features. Figure \ref{fig:bfield_comps} illustrates several different components of the total magnetic field, including the red and cyan features. As discussed, these features were seeded by corks which were selected due to their vicinity to lines with high coronal $\vec{\jmath}/\vec{B}$ values (right panel of Figure \ref{fig:sheet_job}) as well as their vicinity to current sheets (Figure \ref{fig:jz}).

The green field lines in Figure \ref{fig:bfield_comps} represent an ambient overlying horizontal field seeded by 40 random seeds located above 5.7 Mm. This field is not strictly aligned with either x or y but rather a corrugated ensemble of nearly-horizontal field lines and it is likely to be the field with which the red and cyan features reconnect. This is likely because the red, cyan, and green fields run nearly anti-parallel to one another and we see strong current dissipation forming at the interface between these features. 

To explore any photospheric connections that are not already visualized by the red and cyan lines, the orange lines in Figure \ref{fig:bfield_comps} are meant to emphasize the strongest negative-polarity photospheric flux concentrations. This is because the strongest positive-polarity photospheric flux concentrations are mainly represented by the footpoints of the cyan arcade and red flux rope. We note that this polarity alignment confirms that the two features are not reconnecting with each other, but rather the overlying horizontal field. The orange lines were seeded by 30 random seeds with a strong bias towards negative $B_z$. These biases aim to exploit the ways by which the field is connected to the photosphere, connected to the reconnection region, and/or part of the overlying horizontal field which is likely a product of boundary conditions and cross-boundary loops. 

With the magnetic features represented in Figure \ref{fig:bfield_comps}, we can see footpoint concentrations, field lines in the process of reconnecting, and an overlying field with which those lines are reconnecting. All of these features are interconnected; the photospheric flux concentrations move as a result of convective motion, the associated magnetic loops and ropes change as a function of that motion, and the overlying horizontal field reconnects with lines that rise high enough to meet it. Although the red and cyan lines represent the features most relevant to the reconnection region, the photospheric and coronal features are also present and necessary for understanding the overall picture.

\subsubsection{Evolution of the field}
\begin{figure*}
\begin{subfigure}{.5\textwidth}
  \centering
  \includegraphics[width=0.85\linewidth]{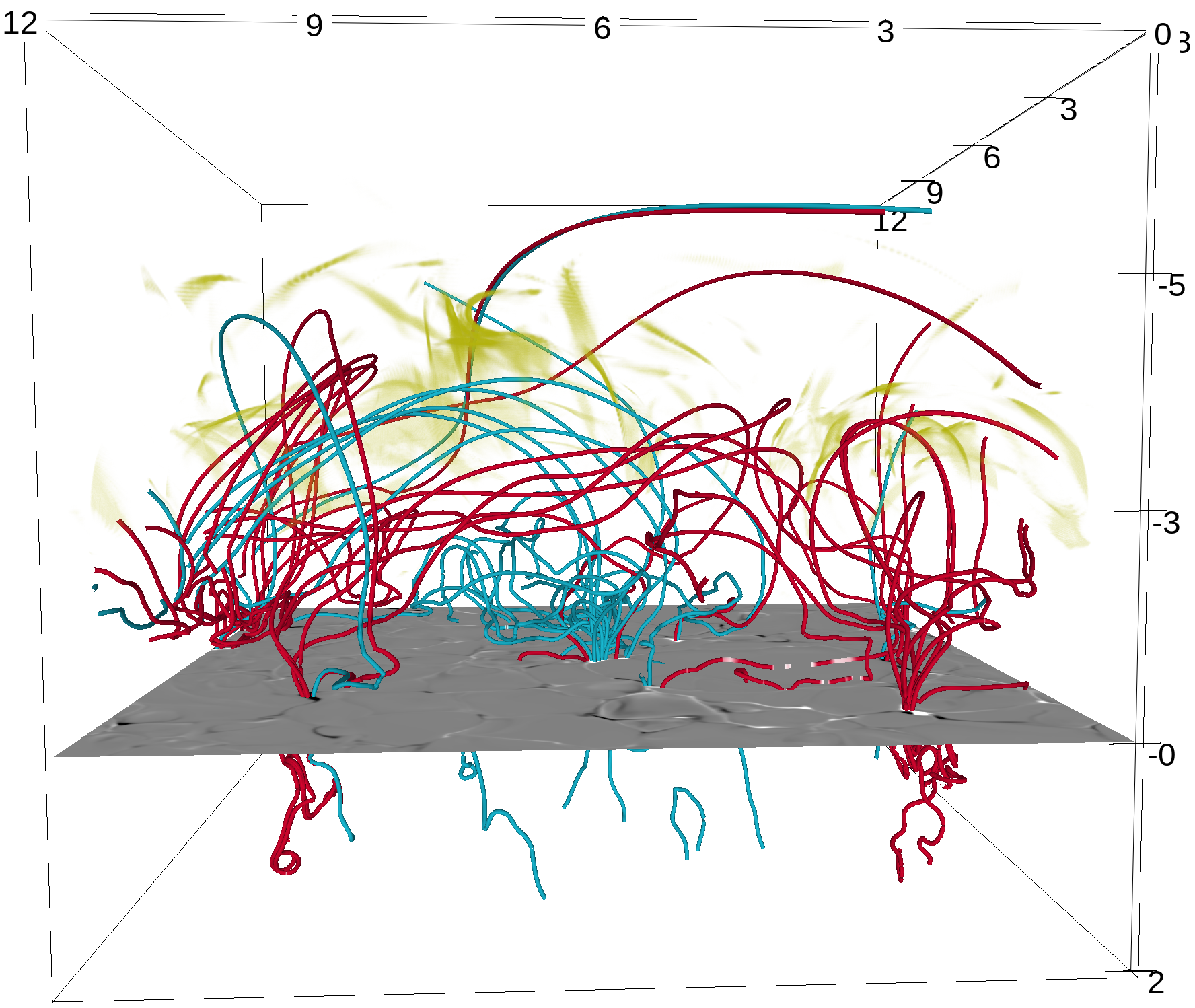}  
  \caption{t = 9 669 s}
  \label{fig:corksa}
\end{subfigure}
\begin{subfigure}{.5\textwidth}
  \centering
  \includegraphics[width=0.85\linewidth]{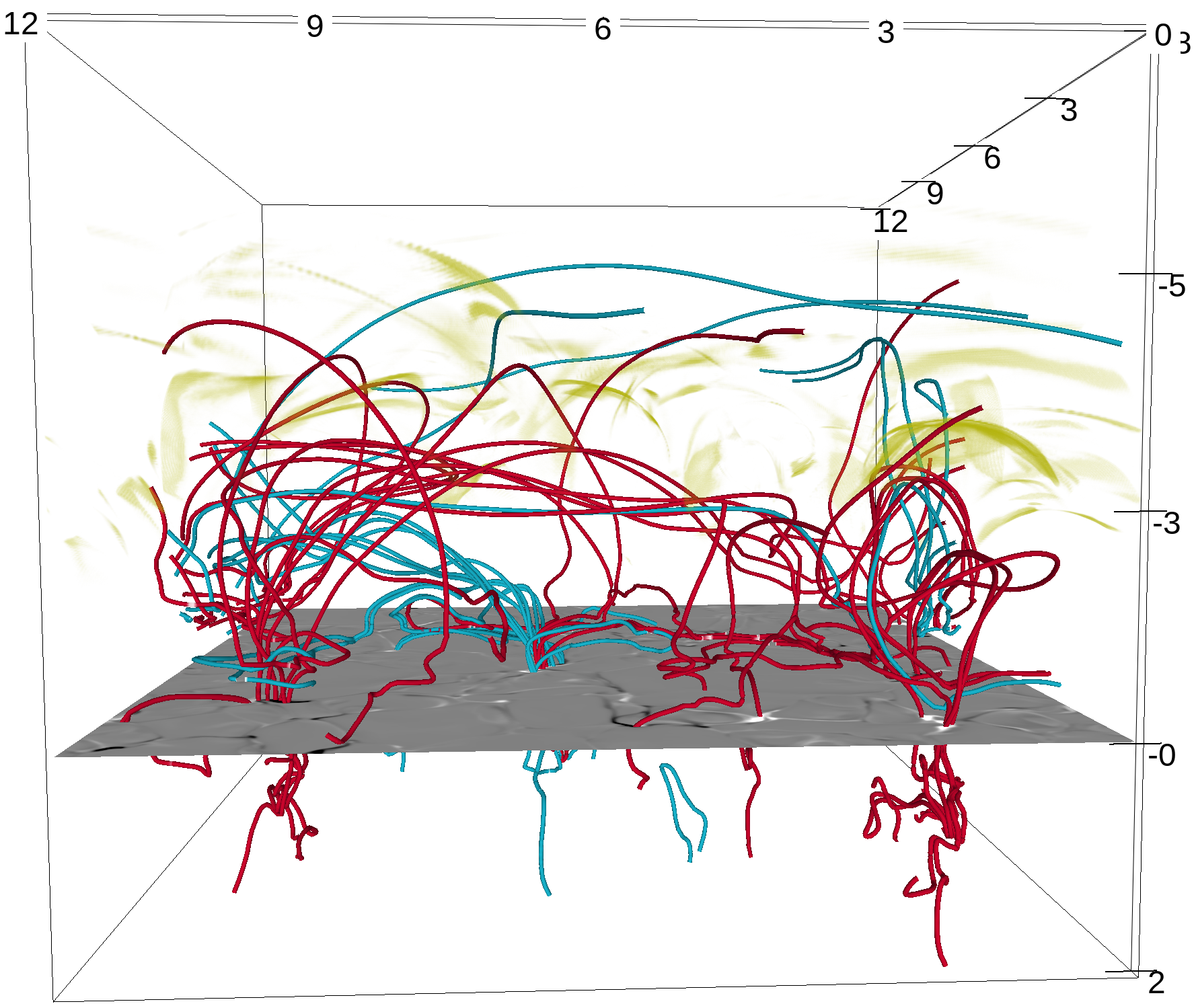}  
  \caption{t = 10 339 s}
  \label{fig:corksb}
\end{subfigure}
\begin{subfigure}{.5\textwidth}
  \centering
  \includegraphics[width=0.85\linewidth]{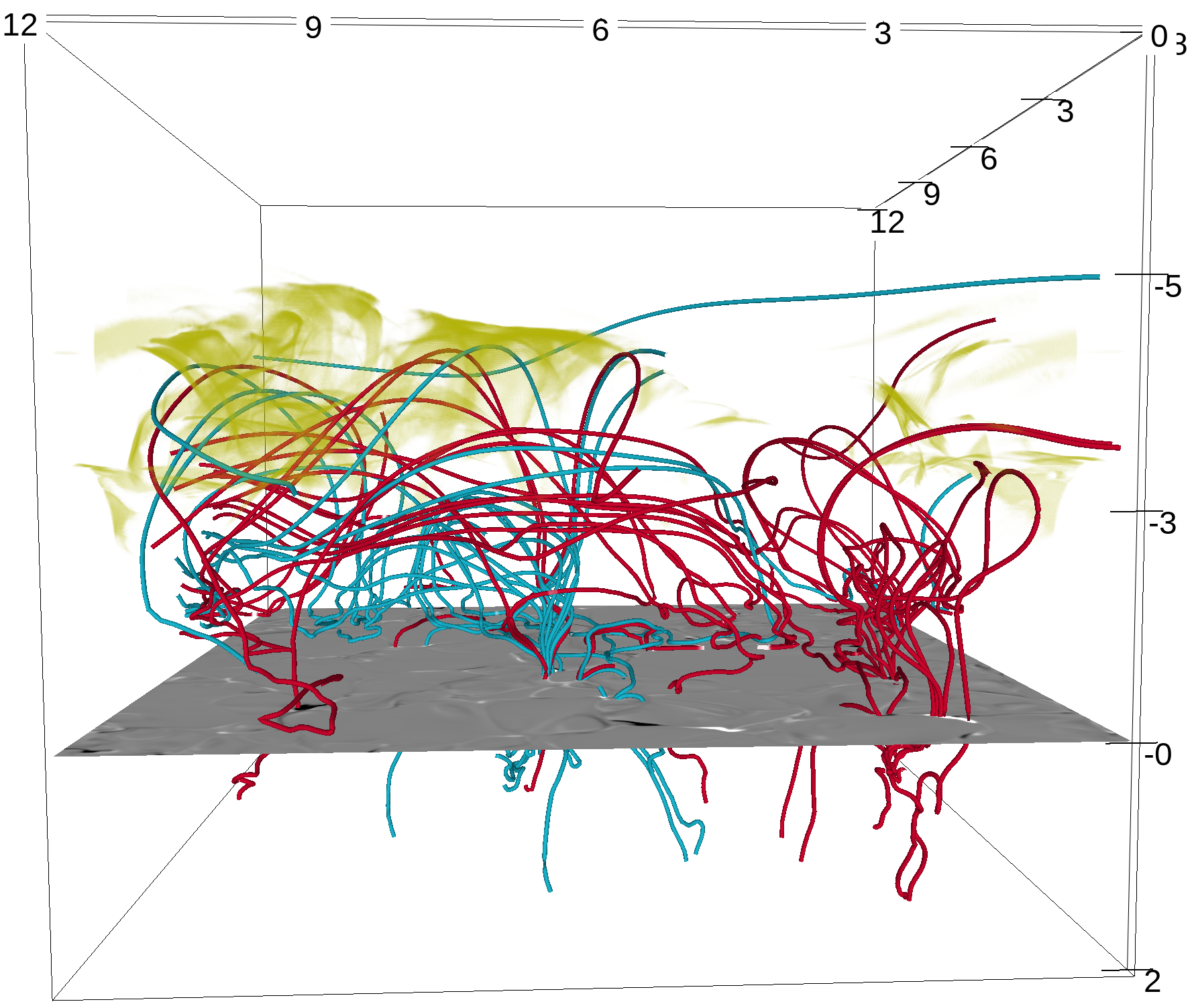}  
  \caption{t = 10 840 s}
  \label{fig:corksc}
\end{subfigure}
\begin{subfigure}{.5\textwidth}
  \centering
  \includegraphics[width=0.85\linewidth]{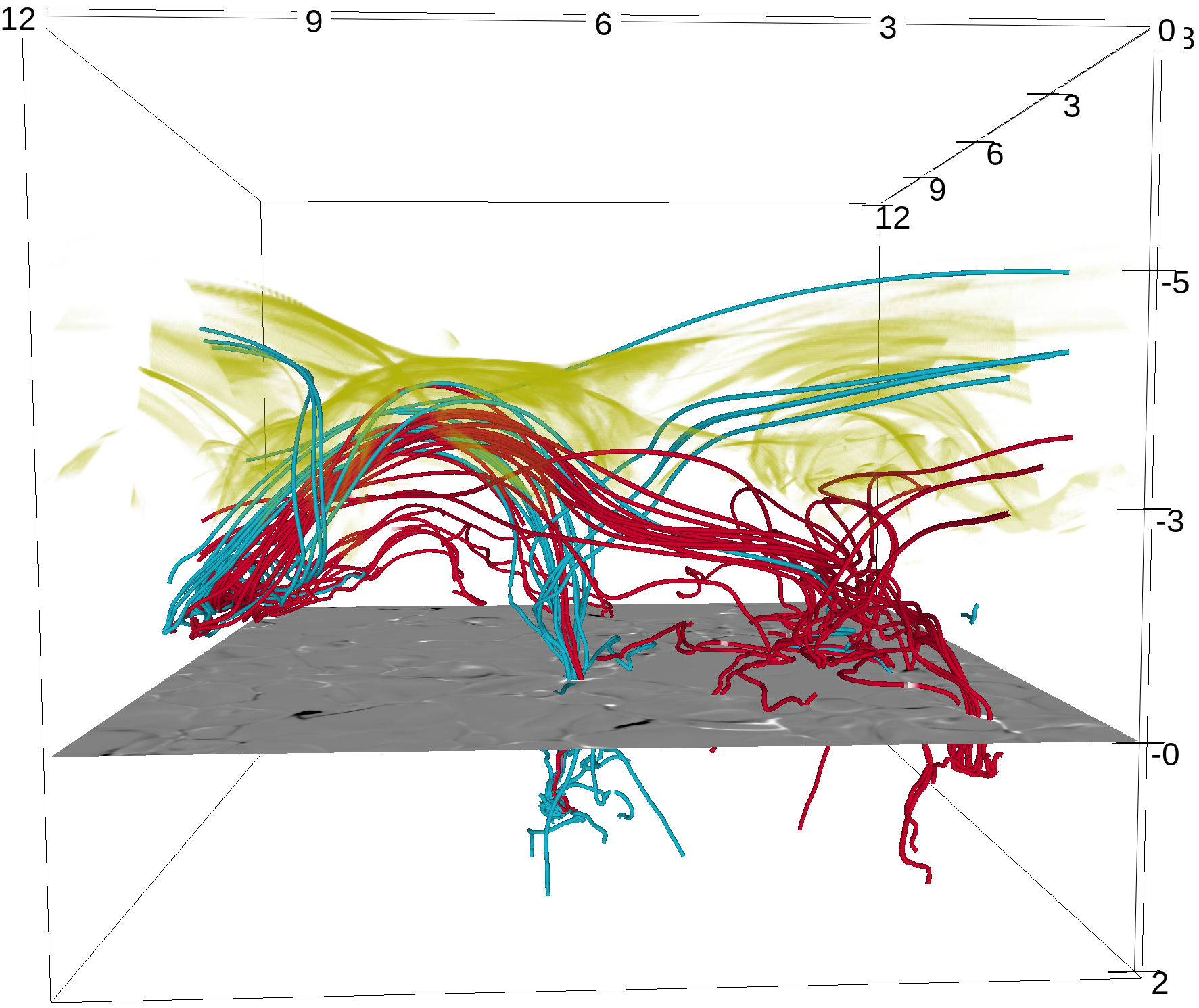}  
  \caption{t = 11 270 s}
  \label{fig:corksd}
\end{subfigure}

\begin{subfigure}{.5\textwidth}
  \centering
  \includegraphics[width=0.85\linewidth]{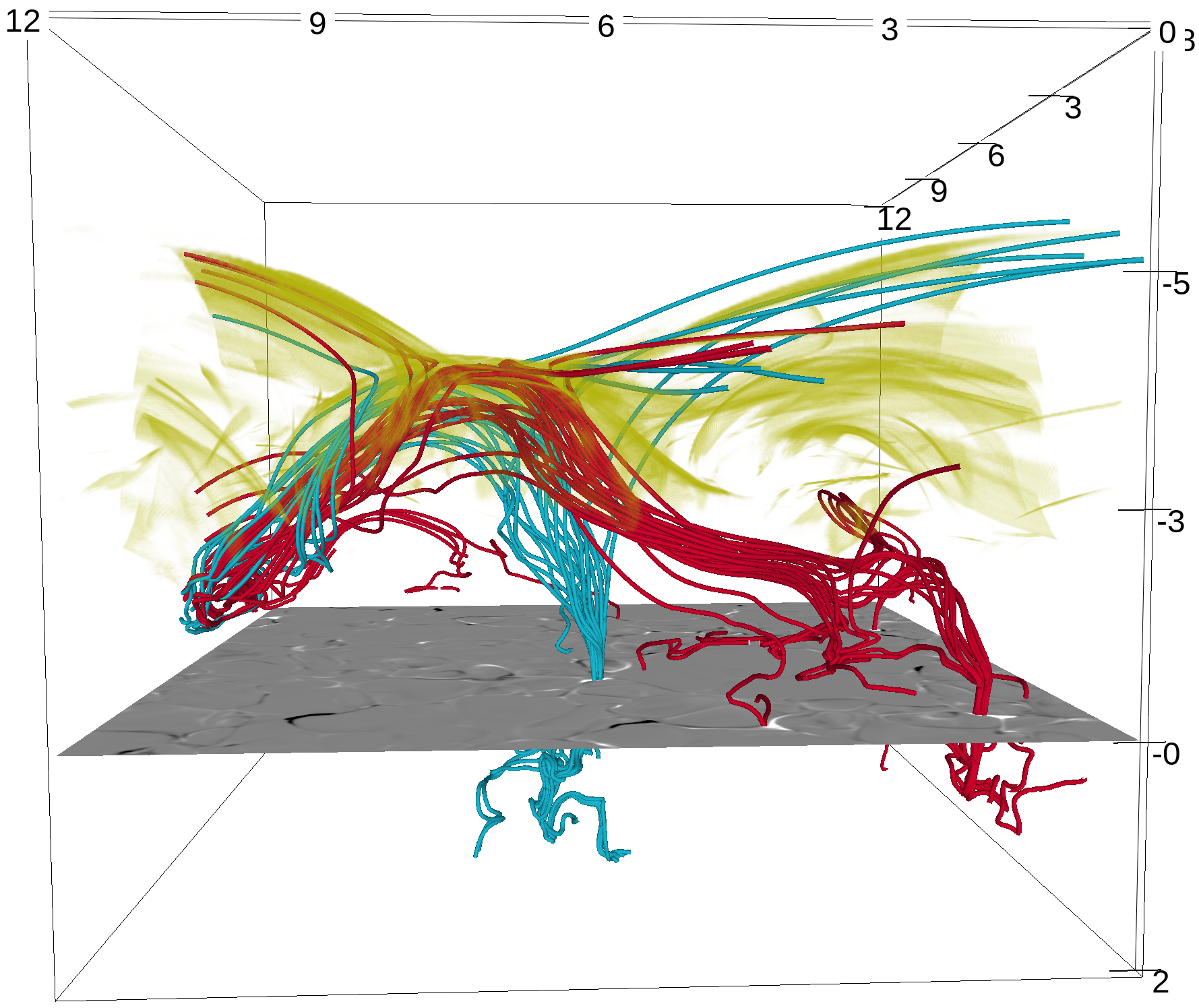}  
  \caption{t = 11 360 s}
  \label{fig:corkse}
\end{subfigure}
\begin{subfigure}{.5\textwidth}
  \centering
  \includegraphics[width=0.85\linewidth]{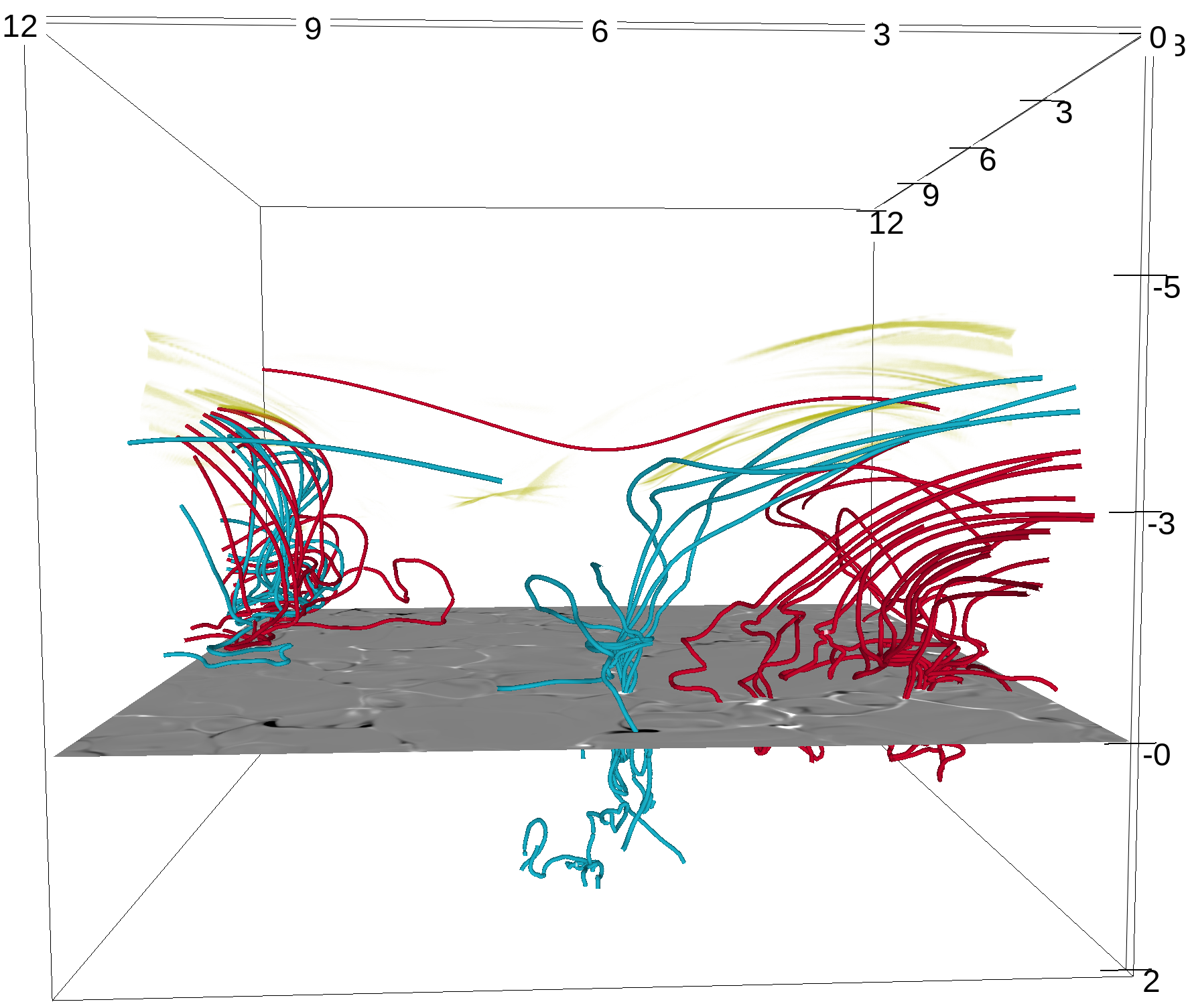}  
  \caption{t = 12 070 s}
  \label{fig:corksf}
\end{subfigure}

\caption{Time series of visualizations of the magnetic field lines seeded and traced by selected corks in the corona. The cyan lines represent an arcade structure while the red lines represent a horizontal flux rope during the time of reconnection. The yellow patches are a volume rendering of Joule heating for each timestamp. A photospheric magnetogram is given for reference.}
\label{fig:corks}
\end{figure*}

Figure \ref{fig:corks} shows a time series of the topological features seeded by the selected corks as discussed above. Recall that these corks trace the evolution of two magnetic features: a horizontal flux rope (red) and an arcade rooted firmly in a central footpoint (cyan). These features do not explicitly include lines that belong to the overlying horizontal field at the time of reconnection, but some of the corks trace the overlying horizontal field at other points in time due to a combination of field evolution and periodic boundary conditions. 

We recall that the horizontal extent of the simulation box is only 12 Mm and the aforementioned features fill nearly the entire cube, with some extending beyond the cube. This implies that the overlying horizontal field physically represents an overlying loop system that is larger than the simulation box and, in practice, this may be a degeneracy of some of the loops that connect across the boundary. It is therefore important to consider the effect of our periodic horizontal boundary conditions, since many of the larger magnetic features are artifacts of each other and must be rooted somewhere in the simulated photosphere via one magnetic field line or another. Here, it is not the goal to trace the full physical extent of each loop across the horizontal boundaries; instead, the aim is to understand how the lines interact with each other within our simulation box as is. 

Each panel of Figure \ref{fig:corks} shows one snapshot of the magnetic features traced by the selected corks, including a volume rendering of the Joule heating and a photospheric magnetogram at that time. For the purposes of visualization, we did not trace the lines across the boundaries even though we have periodic horizontal boundary conditions in the simulation. This is to ensure a clearer picture of the target features, as a periodic tracing of these features simply would add to the overlying horizontal field lines as well as some of the photospheric connections. 

Between t=9\,669 s and t=10\,840 s, Figures \ref{fig:corksa}-\ref{fig:corksc} illustrate the tangled magnetic field lines before they gradually become ordered in preparation for the main reconnection event. While we see no coherent emergence of flux that signals impending reconnection in any obvious way, it is clear that many of the cyan arcade lines are associated with a strong flux concentration near the center of the box, and the red flux rope lines are associated with other flux concentrations near both the right and left boundaries. Other than that, we see no coherent flux emergence during these times.  

Between t=11\,270 s and t=11\,360 s, Figures \ref{fig:corksd} and \ref{fig:corkse} illustrate the magnetic features once the field has ordered into the cyan arcade and red flux rope. At these times, the photospheric connections of the cyan arcade and red flux rope are even more distinct on the right side of the box but the features are entangled over the boundary on the left side. We note that the red flux rope is no longer associated with the flux concentration on the left side of the box as it was in Figures \ref{fig:corksa}-\ref{fig:corksc}. That flux concentration still exists at those later times (see orange lines in Figure \ref{fig:bfield_comps}) but it is no longer associated with this particular flux rope, meaning the feature has changed connectivity over the left boundary. We recall that Figure \ref{fig:corkse} represents the reference snapshot at t = 11\,360 s when the cyan arcade and red flux rope lines are in the process of reconnecting.

During reconnection itself, each line of the cyan arcade and red flux rope joins the overlying horizontal field (not visualized in Figure \ref{fig:corks}, but shown for t = 11\,360 s in Figure \ref{fig:bfield_comps}) and reconnects over the horizontal boundaries. This happens until each line has reconnected, as illustrated in Figure \ref{fig:corksf} at t = 12\,070 s. It is notable that the field ordering, while convoluted in the beginning, more or less remains in the end. Whether a line has been (or will be) a cyan arcade line or a red flux rope line is highly dependent on its photospheric connection on the right side of the box, which is especially evident in the final panel of Figure \ref{fig:corks}. While the two features remain tangled over the left horizontal boundary, the cyan arcade lines are still firmly rooted in the central footpoint and the red flux rope lines remain rooted in photospheric flux concentrations near the right horizontal boundary. 

To compare field line evolution with heating, each panel of Figure \ref{fig:corks} also includes a volume rendering of the Joule heating in 3D (similar to the left panel of Figure \ref{fig:sheet_job}) as they compare to the magnetic features. These snapshots reveal the locations of heat dissipation in the simulation which themselves are associated with the magnetic reconnection. We note that the heat dissipation is most prominent after the arcade and flux rope have ordered themselves as well as at the interface between those features and the overlying horizontal field. It is at that interface where the plasma temperature reaches up to 1.47 MK, resulting from the most energetic atmospheric heating event in this simulation. 

\begin{figure*}
\begin{subfigure}{.5\textwidth}
  \centering
  \includegraphics[width=\linewidth]{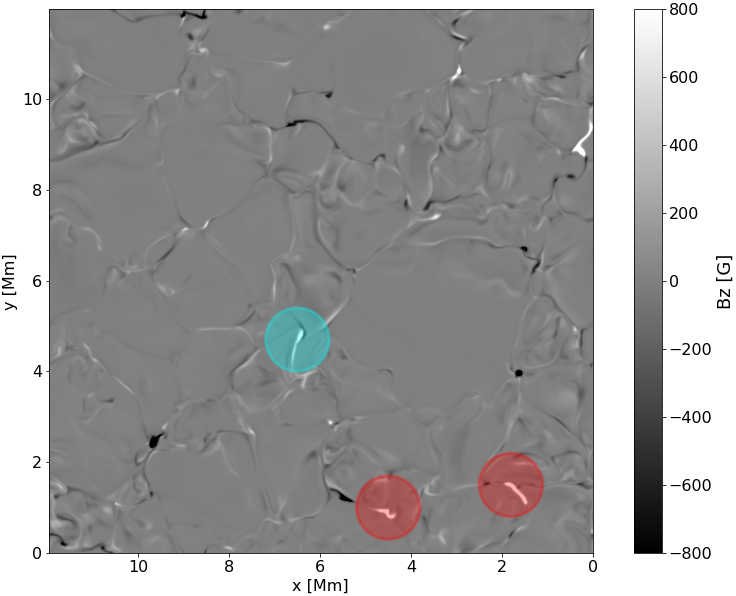}
  \label{fig:bz1240}
\end{subfigure}
\begin{subfigure}{.49\textwidth}
  \centering
  \includegraphics[width=\linewidth]{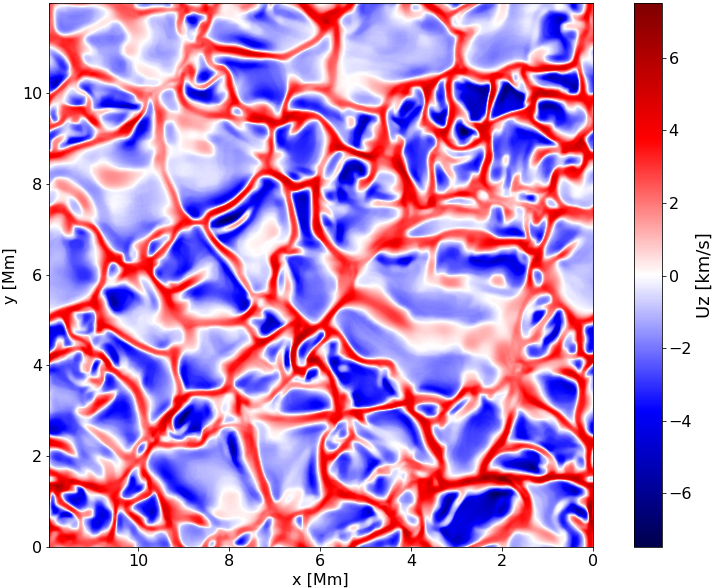}  
  \label{fig:uz1240}
\end{subfigure}
\begin{subfigure}{.5\textwidth}
  \centering
  \includegraphics[width=\linewidth]{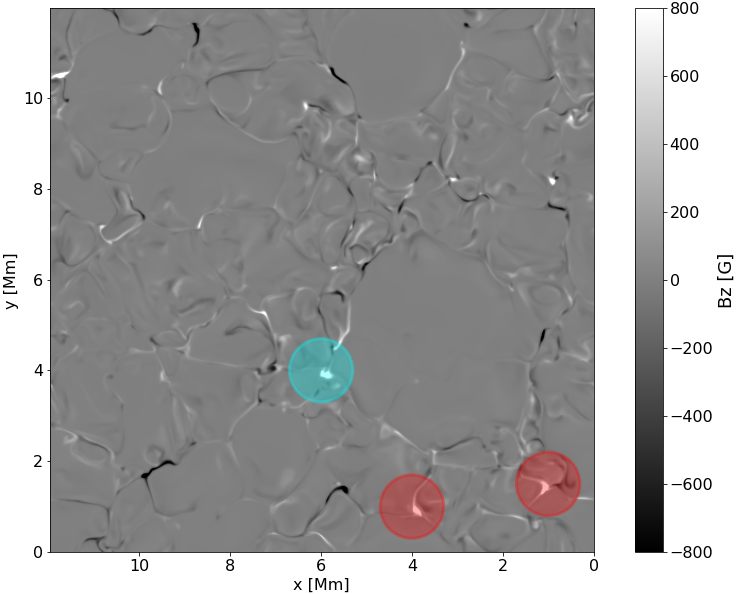}
  \label{fig:bz1300}
\end{subfigure}
\begin{subfigure}{.49\textwidth}
  \centering
  \includegraphics[width=\linewidth]{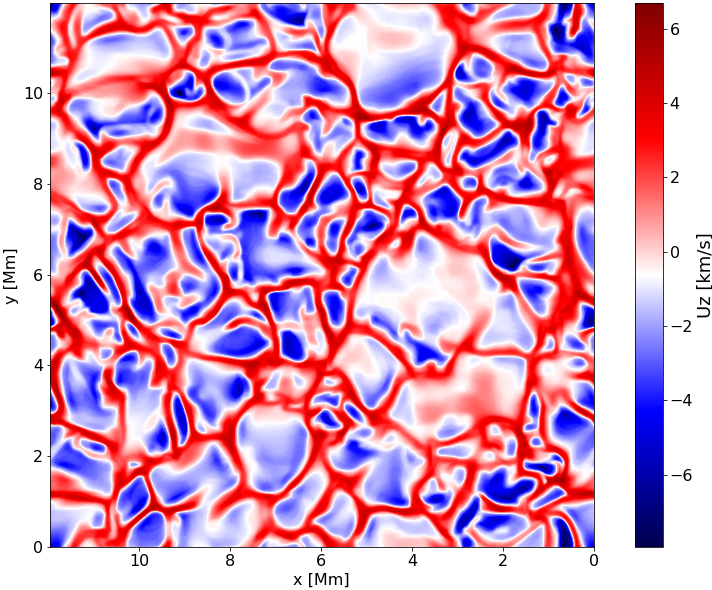}  
  \label{fig:uz1300}
\end{subfigure}
\caption{Photospheric magnetogram $B_z$ (left) and surface velocity $u_z$ (right) for t = 10\,840 s, before the flux rope and arcade are ordered, shown at the top. Photospheric magnetogram $B_z$ (left) and surface velocity $u_z$ (right) for t = 11\,440 s, after the main reconnection event, shown at the bottom. Cyan circles on the magnetograms represent the main positive photospheric root for the arcade, and red circles represent the main positive photospheric roots for the flux rope. The difference in time between the top and bottom panels is 600 s, accounting for roughly one photospheric turnover time.}
\label{fig:cvn}
\end{figure*}

\section{Discussion}\label{dis}

In this work, we provide a case study that supports the idea that convection-driven quiet Sun small-scale magnetic loops can evolve into reconnecting magnetic features that contribute to solar atmospheric heating. 

To trace the time variation of the most relevant magnetic features to reconnection, we selected Lagrangian markers as a function of their proximity to field lines with high coronal values of electric current densities (normalized to the local magnetic field)
(Figure \ref{fig:sheet_job}) and to elongated current sheets (Figure \ref{fig:jz}). The shape of the current sheets in the top panels of Figure \ref{fig:jz} show two joint oval features that indicate the existence of two distinct magnetic flux tubes grazing each other, as shown in the bottom panels. There these are shown as a cyan arcade and as a red horizontal weakly twisted flux rope (twist illustrated in Figure \ref{fig:stream}.) We then followed the evolution of these specific magnetic field lines as well as the flow of the fluid, via the motion of those selected Lagrangian markers. Those allowed us to construct a time series of the evolving arcade and flux rope.

\subsection{Deviations from previous studies}
According to the evolution of the field lines passing through the markers, we cannot define any coherent flux emergence throughout this simulation. No signature common to earlier twisted flux emergence simulations \citep[e.g.,][]{2001ApJ...554L.111F,  2001ApJ...559L..55M, 2004ApJ...610..588M, 2008A&A...492L..35A} can be identified. In addition, we cannot  find any large-scale twisting motion \citep[as modeled in][]{2003A&A...406.1043T, 2005A&A...444..961A, Jiang2021MHDMO}, nor any  systematic and well-aligned series of converging motions and photospheric flux cancellations \citep[e.g.,][]{1989ApJ...343..971V, 1995ApJ...446..377F, 2010ApJ...708..314A, 2011ApJ...742L..27A, 2015ApJ...814..126Z, 2022A&A...659A..25D}. Therefore, despite the pre-reconnecting lines organizing into the aforementioned arcade and flux rope at the onset of reconnection (Figures \ref{fig:corksd} and \ref{fig:corkse}), they do not begin as these coherent structures beforehand (Figures \ref{fig:corksa}-\ref{fig:corksc}). 
Instead, the magnetic field begins as a tangled collection of lines, which gradually order themselves into a coherent flux rope with a neighboring arcade and ultimately reconnect with pre-existing overlying field lines (as illustrated in Figure \ref{fig:bfield_comps}).

Previous studies have also explored the effects of constant magnetoconvective flux emergence on atmospheric processes \citep[e.g.,][]{2017ApJS..229...17S,2018A&A...615L...9C}. In our study, we do not find convincing evidence of new flux emergence that pertains to the main arcade and flux rope features. Figure \ref{fig:cvn} illustrates \textit{$B_z$} and \textit{$u_z$} for two different timestamps. The top panels represent t = 10\,840 s, while the bottom panels represent t = 11\,440 s, a difference of ten minutes. This corresponds to a typical lifetime for a photospheric granule \citep{1961ApJ...134..312B,1978A&A....62..311M,1987A&A...174..275A} and the convective turnover time increases with depth as the pressure scale height also increases with depth into the convection zone  \citep{1996Sci...272.1286C,1998SSRv...85...19C,2003PhDT.......190W,2006astro.ph..6624N,2009LRSP....6....2N,2010A&A...510A..46L}. This typical photospheric turnover period represented in Figure \ref{fig:cvn} is longer than the duration of our reconnection process, beginning before the arcade and flux rope lines have ordered themselves and ending after the main reconnection event. We note that the two highest peaks in Joule heating (Figure \ref{fig:qtavg_cx}) occur between roughly t=11\,200 s and t=11\,500 s, a timescale that is half as long as the timescale represented in Figure \ref{fig:cvn}. While there are certainly small-scale, visible differences in flux between the two magnetograms, there is no large-scale emergence visible in either the magnetogram or the vertical velocity. This indicates no evidence for any relevant new or ``recycled'' flux processes \citep[e.g.][]{2001ASPC..236..363P,2018A&A...615L...9C}. Furthermore, the main positive photospheric roots that anchor the cyan arcade and red flux rope remain present throughout the turnover time as indicated in the left panels of Figure \ref{fig:cvn}, with the cyan and red circles respective to their corresponding magnetic feature. While the photospheric roots certainly move, change, and interact with other smaller concentrations throughout the simulation, they remain strong before, during, and after the main reconnection events. This suggests that the atmospheric processes responsible for ordering the coronal lines into an arcade and flux rope are happening in addition to background magnetoconvective processes. 

\subsection{Similarities to idealized models}
Since the convection-driven photospheric field is constantly moving and changing, coronal field lines are always reconnecting to some extent throughout the simulation via flux-braiding \citep[as in, e.g.,][]{2005ApJ...618.1031G, 2005ApJ...618.1020G,  2015ApJ...815....8R, 2015RSPTA.37340265W}. The arcade and flux rope, although they are the topological features most relevant to the largest reconnection events in the simulation, are therefore not the only reconnecting lines. The reconnection of other atmospheric field lines in the corona as well as at their footpoints are likely contributors to the ordering of the arcade and flux rope lines, up to the onset of the most energetic reconnection event.

The effects of large-scale photospheric footpoint motion on null-less magnetic reconnection in 3D have already been extensively studied by various groups \citep[e.g., ][]{2003ApJ...595..506G, 2005A&A...444..961A, 2006SoPh..238..347A, 2006SSRv..124..345B, 2016SoPh..291..143E} All show that many types of footpoint motions can lead to the expansion of magnetic loops to the point of eventual reconnection \citep[in line with 2D models from, e.g.,][]{1988ApJ...324..574L}. In all of these idealized experiments, a set of bipolar features with four footpoints was prescribed and forced with varying footpoint motion until reconnection occurred in the simulated corona. 

In these models, there is no convection and, therefore, no self-consistent motion; the idealized footpoint motion is the prescribed driving force. Their setup included no dynamic chromosphere, a low corona dominated by the Lorentz force, and line-tied conditions at the photosphere which make the photospheric boundary infinitely conductive, inertial, and reflective \citep[see][for detailed explanations]{2005A&A...430.1067A}. With that, it was fairly straightforward to follow the exact field evolution in time and follow the connections between the photosphere and corona. This is far more complicated in our simulation, where convective motion is the source of photospheric flux concentrations and our realistic chromosphere muddles the relationship between the photosphere and corona. Even so, the fact that we see similar behaviors as in, for instance,  \citet{2005A&A...444..961A} suggests that the same fundamental physics applies in both models.

The models appear to validate each other, at least in terms of the spatiotemporal behavior of the current sheet formation and null-less reconnection. Since we see similar behavior in both idealized (i.e., zero-$\beta$ and line-tied) and more complex (i.e., fully stratified and convective) cases, it is likely that the results discussed in \citet{2005A&A...444..961A} represent the backbone of what is happening in our simulation. Further, we see that neither photospheric dynamics driven by local convective motion nor a realistic chromosphere destroy the possibility of seeing a hyperbolic flux tube (HFT) with magnetic reconnection in the corona. Our results also support the assertion in \citet{2005A&A...444..961A} that in the presence of a HFT, photospheric footpoint motions  -- whatever their form -- can result in current sheet formation at the HFT and the subsequent onset of reconnection. Therefore, even with varying levels of complexity, the models still share the same general physical principles. 

Footpoint motion may explain some of the evolution of the arcade and flux rope, especially since these features tend to display a preference for specific photospheric flux concentrations as roots. Although the two features are entangled with one another over the left periodic horizontal boundary (see, e.g., Figures \ref{fig:bfield_comps} and \ref{fig:corks}), they both have distinct positive photospheric roots near the right horizontal boundary as especially seen in Figure \ref{fig:corksf}. While the positive photospheric roots do evolve and sometimes coalesce with other smaller flux concentrations, they remain throughout the turnover time of our simulated convection cells as illustrated in Figure \ref{fig:cvn}. This implies that these roots are not built up by new flux emergence or flux recycling, but rather, they remain steady long enough for the two magnetic features to form in the atmosphere. This may suggest that features like these do not need to emerge coherently in order to form; they simply need a reliable network of long-lasting and relatively strong photospheric connections to root them. This conclusion could provide a context for observations of magnetic reconnection that are not related to obvious coherent flux emergence. 

\subsection{Building on idealized models}
As mentioned earlier in this paper, there are more reconnection events occurring in the box than just reconnection near the footpoints. Multiple atmospheric reconnections may also play a role in ordering the arcade and flux rope features before they eventually reconnect with the overlying horizontal field and produce major heating events. During the simulation, lines and loop systems that are located above the initially-disorganized arcade and flux rope lines can also reconnect with the overlying horizontal field earlier. This can effectively remove some of those lines from the immediate atmosphere above, as well as reconnect them across the boundaries. In addition, these earlier reconnections may bring down some horizontal flux to feed the flux rope formation. With that, the arcade and flux rope have ample space and time to build up until they eventually reconnect with the overlying horizontal field themselves. 

Even with sufficient time and space to form, the smooth ordering of a coherent twisted flux rope from an initially tangled field can be puzzling. We note that the arcade and flux rope are formed from lines that already exist in the corona rather than new flux emerging from the photosphere. These are lines that consistently pass through the same Lagrangian markers throughout the simulation, meaning that they are already extended into the corona from t=9\,669 s. We argue that it may be the result of the ``inverse cascade'' of magnetic helicity \citep[as proposed by][]{1974PhRvL..33.1139T,1975JFM....68..769F}. More recently, the redistribution of helicity at large scales via multiple reconnections has been studied by \citet{2015ApJ...805...61Z}, \citet{2017ApJ...835...85K}, and \citet{2019ApJ...883..148R}, among others. In these studies, systems of plane-parallel loops-in-a-box are given different driving forces at their photospheric boundaries, some of which encourage reconnection of the twisting field lines. Under certain driving forces, the conditions for small-scale reconnection are met and upon reconnection, the helicity of the loops is redistributed to larger scales rather than smaller scales. These 3D numerical realizations of the inverse helicity cascade could actually describe the same process responsible for the ordering of the arcade and flux rope lines in our complex and fully-stratified simulation. Within our box of dynamic plasma it is challenging to separate one cause-and-effect from another, but not impossible. Our self-consistent driver reveals several pre-reconnection features likely formed by inverse helicity cascade, one example of which is illustrated by the large-scale magnetic twist in Figure \ref{fig:stream}. This large-scale twist is not present throughout the entire simulation to the extent that it is shown in Figure \ref{fig:stream}, but rather, it forms by the coalescence of several smaller-scale helical features. First, the twist builds up until it reaches the scale of the connectivity domain, then reconnection occurs. While inverse helicity cascade should theoretically produce helical features on the scale of the computational box, this is impossible in our case because the arcade and flux rope eventually meet an overlying horizontal field and undergo comparatively large-scale reconnection. Further details are beyond the scope of this paper, and the process of inverse helicity cascade is an idea that will be further explored in a future study.
 
\subsection{Link between reconnection and heating}
Finally, reconnecting field lines are easily isolated by looking for high values of $\vec{\jmath}/\vec{B}$ at any timestep. Following this, it is simple to determine where the biggest reconnection events are located throughout the duration of the simulation. Although reconnection is constant, atmospheric heating to 1.47 MK is not constant. It follows that quiet Sun heating events are only as significant when the corresponding magnetic features have had time to store sufficient magnetic energy to release as kinetic and thermal energy upon reconnection, as with the arcade and flux rope. In this case, the arcade and flux rope were attached to long-lasting photospheric flux concentrations on at least one side, shielded by loops that join the overlying horizontal field first, and able to build up twist as they interacted with each other at the onset of their reconnection. Under these conditions, we see that this convection-driven field configuration is capable of generating atmospheric temperatures up to 1.47 MK. It is not unreasonable to expect such conditions on the real Sun, where convection-driven magnetic features and intertwining coronal loop systems exist. Despite the impulsive nature of our simulated events, the sum of several comparable events could contribute to overall atmospheric heating above quiet Sun photospheres.

\section{Conclusions}\label{con}
Our convection-driven \textit{Bifrost} simulation of the quiet Sun is demonstrably capable of producing impulsive atmospheric heating events which generate plasma temperatures up to 1.47 MK and Joule heating on the order of 5.4 $\times 10^{17}$ J. The largest heating event was a result of two key magnetic features, an arcade and a horizontal weakly-twisted flux rope, which both reconnect with an overlying horizontal field after sufficient time to build up and order.

In contrast to previous idealized studies where photospheric drivers are purposefully prescribed with relatively smooth distributions, our photospheric drivers self-consistently result from solar convection. Furthermore, there is little evidence that the magnetic features relevant to reconnection are associated with coherent flux emergence, twisting motions, or flux cancellation. Instead, they rely more heavily on their photospheric connections on one side. There is evidence that ambient reconnection events should contribute to the ordering of the magnetic field before the onset of the main reconnection event, suggesting a possible inverse cascade of helicity. 

Reconnection occurs throughout the simulation, but our most energetic reconnection event requires sufficient time to order the field and store enough magnetic energy. We assert that similar reconnection events are likely in the real Sun and this case study may shine some light on the ways by which the quiet Sun contributes to coronal heating. 

Further studies are required to understand the specifics of helicity cascade as it contributes to ordering the magnetic field as well as the onset of the largest reconnection event. These specifics will be explored in a subsequent paper. Furthermore, the synthesized observational signatures of such heating events may help us understand what these events look like in real observations. The synthesis of chromospheric and coronal EUV lines will be compared to IRIS, SDO, and EUI observations. These comparisons may prove useful for understanding the signatures of quiet Sun magnetic reconnection that is not associated with coherent flux emergence, but that is  driven, rather, entirely by convective motion.

\begin{acknowledgements}
      This research was supported by the Research Council of Norway through its Centres of Excellence scheme, project number 262622, and through grants of computing time from the Programme for Supercomputing. We acknowledge funding support by the European Research Council under ERC Syngergy grant agreement No. 810218 (Whole Sun). We also acknowledge the use of Visualization and Analysis Platform for Ocean, Atmosphere, and Solar Researchers (VAPOR version 3.6.1) for field tracing and heat rendering figures. VAPOR is a product of the Computational Information Systems Laboratory at the National Center for Atmospheric Research. R.A.R. acknowledges the insightful and motivating conversations with Pascal D\'{e}moulin and \'{E}tienne Pariat on the matter of inverse helicity cascade. G.A. acknowledges financial support from the French national space agency (CNES), as well as from the Programme National Soleil Terre (PNST) of the CNRS/INSU also co-funded by CNES and CEA. G.A. also thanks the whole RoCS group for their warm hospitality during his stays in Oslo.
\end{acknowledgements}

\bibliography{biblio}

\end{document}